\documentclass[11pt,citesort]{article}

\usepackage{epsfig}
\usepackage{cite}
\usepackage{amsfonts,color}
\usepackage{hyperref}
\usepackage[centertags]{amsmath}     
\usepackage{amssymb,amscd}
\usepackage{cancel}

\newcommand{\nn}{\nonumber}
\newcommand{\gzero}{g_0^\theta}
\newcommand{\gone}{g_1^\theta}
\newcommand{\mn}{m_N}
\newcommand{\mpi}{M_{\pi}}
\newcommand{\fpi}{F_{\pi}}
\newcommand{\NfourLO}{{\sf N$^4$LO}}
\newcommand{\NthreeLO}{{\sf N$^3$LO}}
\newcommand{\NtwoLO}{{\sf N$^2$LO}}
\newcommand{\NLO}{{\sf NLO}}
\newcommand{\LO}{{\sf LO}}
\newcommand{\he}{{}^3{\rm He}}
\newcommand{\hy}{{}^3{\rm H}}
\newcommand{\CPv}{\cancel{\rm CP}}

\begin{document}
\begin{titlepage}

\vspace{2.0cm}

\begin{center}
{\Large\bf Nuclear Electric Dipole Moments\\[1mm] in Chiral Effective Field Theory}
\vspace{1.2cm}

{\large \bf J. Bsaisou$^a$, J. de Vries$^a$, C. Hanhart$^{a,b}$, S. Liebig$^a$,\\[1mm] Ulf-G. Mei{\ss}ner$^{a,b,c,d}$, D. Minossi$^{a}$, A. Nogga$^{a,b}$,
and A. Wirzba$^{a,b}$} 

\vspace{0.5cm}

{\large 
$^a$ 
{\it Institute for Advanced Simulation, Institut f\"ur Kernphysik, 
and J\"ulich Center for Hadron Physics, Forschungszentrum J\"ulich, 
D-52425 J\"ulich, Germany}}

\vspace{0.25cm}
{\large
$^b$ \it{JARA -- Forces and Matter Experiments, Forschungszentrum J\"ulich, D-52425 J\"ulich, Germany}}

\vspace{0.25cm}
{\large
$^c$ \it{JARA -- High Performance Computing, Forschungszentrum J\"ulich, D-52425 J\"ulich, Germany}}

\vspace{0.25cm}
{\large 
$^d$ 
{\it Helmholtz-Institut f\"ur Strahlen- und Kernphysik and Bethe Center for 
Theoretical Physics, Universit\"at Bonn, D-53115 Bonn, Germany}}

\end{center}

\vspace{1.5cm}

\begin{abstract}
We provide a consistent and complete calculation of the electric dipole moments of the deuteron, helion, and triton in the framework of chiral effective field theory. The CP-conserving and CP-violating interactions are treated on equal footing and we consider CP-violating one-, two-, and three-nucleon operators up to next-to-leading-order in the chiral power counting. In particular, we calculate for the first time EDM contributions induced by the CP-violating three-pion operator. We find that effects of CP-violating nucleon-nucleon contact interactions are larger than those found in previous studies based on phenomenological models for the CP-conserving nucleon-nucleon interactions. 
Our results which apply to any model of CP violation in the hadronic sector 
can be used to test various scenarios of CP violation. As examples, we study the implications of our results on the QCD $\theta$-term and the minimal left-right \mbox{symmetric} model. 
\end{abstract}

\vfill
\end{titlepage}

\setcounter{equation}{0}
\section{Introduction}

Any measurement of a non-vanishing  permanent electric dipole moment (EDM) -- be it  for an electron,  nucleon, nucleus, atom or 
polar molecule with a non-degenerate ground state -- would signal the simultaneous violation of parity (P) and time-reversal (T) symmetry and hence the violation of CP symmetry. The complex phase of the Cabibbo-Kobayashi-Maskawa (CKM) matrix
of the Standard Model (SM) generates EDMs orders of magnitude 
smaller~\cite{Czarnecki:1997bu,Pospelov_review,Mannel:2012qk,Mannel:2012hb} 
than the sensitivities of current and planned experiments. 
Therefore, EDMs serve as ideal probes for {\em flavor-diagonal} CP violation -- with a minimal SM background -- from {\it e.g.} the $\theta$-term of Quantum Chromodynamics (QCD) \cite{'tHooft:1976up} and 
{\em beyond-the-SM} (BSM) physics.
 Popular examples of the latter are, {\it e.g.}, supersymmetric, multi-Higgs, or left-right symmetric models. Irrespectively of the high-energy details of such SM extensions, 
when evolved down to an energy scale 
where QCD becomes
non-perturbative, they give rise to several effective operators of mass dimension six. They are known as the quark EDM (qEDM), quark chromo-EDM (qCEDM), gluon chromo-EDM (gCEDM), and various four-quark interactions \cite{Buchmuller:1982ye,Grzadkowski:2010es,deVries:2012ab}.

Although {\em one} successful measurement of a non-vanishing EDM would already prove the existence of CP violation beyond the CKM-matrix, it would not be sufficient to reveal the underlying source(s) of CP violation. Independent EDM measurements of single nucleons (neutron and proton) and light nuclei, \textit{e.g.} the deuteron, the helium-3 nucleus (helion) and, maybe, the hydrogen-3 nucleus (triton), and heavier systems such as various atoms and molecules, are in general required to learn more about the underlying source(s). The concept of probing the QCD $\theta$-term and BSM physics using EDMs of light nuclei
has attracted much attention in 
recent years~\cite{deVries:2012ab,khriplovich,Pospelov_deuteron,LiuTimmermans,Stetcu:2008vt,Afnan,deVries2011a,deVries2011b,Bsaisou:2012rg,Song:2012yh,dissertation,Dekens:2014jka,Wirzba:2014mka} and is the basic idea underlying plans for EDM measurements in dedicated storage 
rings~\cite{Semertzidis:2003iq,Semertzidis:2011qv,Lehrach,Pretz:2013us,Rathmann:2013rqa}. 
The main advantage of 
{\em light} nuclei is that the associated nuclear physics is theoretically well under control, such that these systems can be used to probe the underlying CP-violating mechanism.

The various sources of CP violation at the energy scale $\Lambda_{\chi}\sim 1$ GeV induce, in principle, an infinite set of CP-violating terms in the effective low-energy pion-nucleon Lagrangian that, however, can be ordered by a 
power-counting 
scheme~\cite{deVries:2012ab,deVries2011b,Liebig:2010ki,Bsaisou:2012rg, dissertation}. It was concluded that the leading EDM contributions for nucleons and light nuclei can be expressed in terms of seven interactions:
\begin{eqnarray}\label{eq:impcoup}
  \mathcal{L}_{\CPv}^{\pi N}
  &=&-\,d_nN^{\dagger}(1-\tau^3)S^{\mu}v^{\nu}NF_{\mu\nu}-d_pN^{\dagger}(1+\tau^3)S^{\mu}v^{\nu}NF_{\mu\nu}\nn\\
  &&\mbox{}+(\mn\Delta)\,\pi_3\pi^2+g_0N^{\dagger}\vec{\pi}\cdot\vec{\tau}\,N+g_1N^{\dagger}\pi_3N\nonumber\\
  &&\mbox{}+\,C_1N^{\dagger}N\,\mathcal{D}_{\mu}(N^{\dagger}S^{\mu}N)+\,C_2N^{\dagger}\vec{\tau}\,N\cdot\mathcal{D}_{\mu}(N^{\dagger}\vec{\tau}\,S^{\mu}N)\,\, .
\end{eqnarray}
Here, $v^{\mu}=(1,\vec{0})$ and $S^{\mu}=(0,\vec{\sigma}/2)$ are the nucleon velocity and spin, respectively, $\vec{\tau}$ denotes the vector of the isospin Pauli-matrices $\tau^{i}$, $\vec \pi =(\pi_1,\pi_2,\pi_3)^T$ the pion isospin triplet,  $\mathcal{D}_{\mu}$ the covariant derivative acting on the nucleon doublet $N= (p\,,n)^T$, $\mn=938.92$ MeV the average nucleon mass \cite{pdg}, and $F_{\mu\nu}$ the electromagnetic field strength tensor. For further notations, we refer to Ref.~\cite{Bernard:1995dp}. 
The first two interactions in Eq.~\eqref{eq:impcoup} are the neutron ($d_n$) and proton EDM ($d_p$), respectively, which are treated as effective parameters here. The second line of Eq.~\eqref{eq:impcoup} contains a purely pionic interaction (with coupling constant $\Delta$) and two pion-nucleon interactions (with coefficients $g_{0,1}$),\footnote{$\Delta$, $g_0$ and $g_1$ are dimensionless and  have the opposite signs of the corresponding dimensionful quantities specified
in \cite{deVries2011b,deVries:2012ab};  $\bar \Delta  =- 2 \fpi \mn \Delta$ 
 of Ref.~\cite{Dekens:2014jka} carries dimensions with
 $\fpi=92.2\,\rm{MeV}$  the pion decay constant~\cite{pdg}.}  while the interactions in the last line denote two CP-violating nucleon-nucleon contact terms.
Other hadronic interactions, such as the  isotensor pion-nucleon interaction $g_2 N^\dagger \pi_3\tau_3 N$ only appear at  orders higher than those considered here.

The different sources of CP violation ({\it e.g.} the $\theta$-term and dimension-six sources) are expected to contribute to all CP-violating operators in Eq.\,(\ref{eq:impcoup}), but at different strengths based on the field content and chiral-symmetry properties of the source \cite{Pospelov_deuteron,deVries2011b,Bsaisou:2012rg,dissertation,Dekens:2014jka}.  Different sources therefore yield different hierarchies of nucleon and nuclear EDM contributions which can explicitly be probed by EDM measurements.

 The main goal of this paper is to provide 
 the results of a {\em complete and consistent} calculation within {\em chiral effective field theory} ($\chi$EFT)\footnote{The extension of
chiral perturbation theory (ChPT) to systems with more than one nucleon.} of the leading single-nucleon, two-nucleon ($2N$) and three-nucleon ($3N$)  contributions to the EDMs of light nuclei up-to-and-including next-to-leading order (\NLO) with {\em well defined uncertainties}. The results are expressed as functions of the seven low-energy constants (LECs) in Eq.\,(\ref{eq:impcoup}), which have to be extracted 
-- in the future -- from a combination of EDM measurements and, whenever possible, supplemented with Lattice-QCD calculations.

This paper is organized as follows: the relevant CP-violating operators yielding leading-order (\LO) and next-to-leading-order (\NLO) EDM contributions for any of the considered sources of CP violation are presented in Section~\ref{sec:edff}, 
 while the employed power-counting scheme  is briefly explained in Appendix~\ref{app:pc}. 
 The EDMs of the deuteron, helion and triton as functions of the coefficients in Eq.\,(\ref{eq:impcoup}) up-to-and-including \NLO\ are computed in Section~\ref{sec:nucedms}, where
$\chi$EFT  as well as phenomenological potentials are employed. The main results are presented in Tables \ref{tab:deuteron} and \ref{tab:3n} 
and Eqs.\,(\ref{eq:h2edm})--(\ref{eq:h3edm}). 
As an application of our result, we discuss the cases of the $\theta$-term and minimal left-right symmetric models in Sections~\ref{sec:theta} and \ref{sec:LR}, respectively, for which the EDMs of the two- and three-body nuclei can be expressed as functions of a single parameter. We conclude this paper with a brief summary and discussion. Appendix~\ref{app:beta} provides
information about the regulator dependence of the EDM contributions resulting from the  short-range $4N$ vertices.  The small
$D$-wave corrections to the single neutron and proton contributions of the
deuteron EDM are missing in the original text. The places where this occurs are marked by
a footnote referring to
 Appendix~\ref{app:erratum} which includes
an erratum to the published version of the paper.

\setcounter{equation}{0}
\section{CP-violating nuclear operators}
\label{sec:edff}

\begin{figure}
\centering
\includegraphics[scale=0.7]{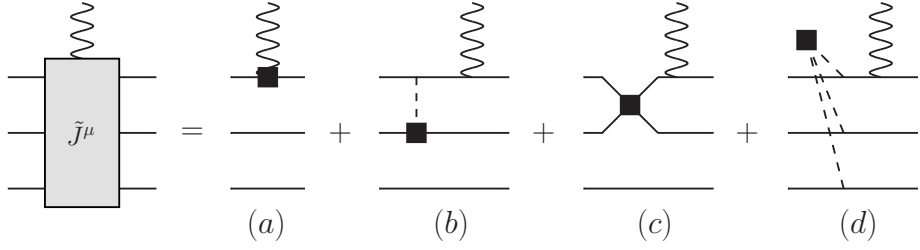}
\caption{Leading contributions to the total CP-violating $3N$ current. A CP-violating vertex is depicted by a black box. Full, dashed, and wiggly lines refer to
nucleons, pions, and photons, respectively. Only one ordering per diagram class is shown. CP-conserving interactions in the intermediate states of diagrams (b)-(d) are not displayed explicitly. When the lowest nucleon line is removed, the diagrams (a)-(c) define the leading contributions for the $2N$ system, too.\label{Fig:current}}
\end{figure}
The electric dipole form factor $F_3^{A}$ for a nucleus $A\!=\!{}^2{\rm H},\,{}^3{\rm He},\,{}^3{\rm H}$ is defined by the nuclear matrix element of the {\em total} CP-violating transition current $\tilde{J}^{\mu}$. Since CP violation is an extremely small effect, only operators with exactly one insertion of a vertex from Eq.\,(\ref{eq:impcoup}) need to be considered. The total CP-violating transition current can be written in short-hand notation as
\begin{equation}\label{eq:ptvcurrent}
\tilde{J}^{\mu}=J^{\mu}_{\CPv}+V_{\CPv}\,G\,J^{\mu}+J^{\mu}\,G\,V_{\CPv}+\cdots\,\,,
\end{equation}
where $J^{\mu}$ ($J^{\mu}_{\CPv}$) denotes the CP-conserving (CP-violating) irreducible transition current, $V_{\CPv}$ the CP-violating potential, and 
$G$ the complete CP-conserving $2N$ or $3N$ propagator. All operators appearing in Eq.~\eqref{eq:ptvcurrent} are calculated consistently within $\chi$EFT. 

The EDM of a nucleus $A$ is most conveniently computed in the Breit-frame, in which the outgoing photon four-momentum equals $q^\mu\!=\!(0,\,\vec{q})$ and $\vec{q}$ can be chosen to point in the $z$-direction, {\it i.e.} $\vec q= (0,\,0,\,q)$. The CP-violating form factor
$F_3^A(q^2)$  and the EDM $d_A$ of a nucleus $A$ are then given by the following matrix element and its $q^2\rightarrow 0$ limit, respectively:
\begin{equation}
-iq\frac{F_3^{A}(q^2)}{2m_A}=\left \langle A; M_J\!=\!J \left |\,\tilde{J}^{0}(q)\,\right |A;M_J\!=\!J \right\rangle\,\, , \qquad d_A
=\lim_{q^2\to 0}\frac{F_3^{A}(q^2)}{2m_A}\,\, .
 \label{eq:FFdef}
\end{equation}
Here, $J$ is the total angular momentum of the nucleus of mass $m_A$ and $M_J$ is its $z$-component.

The nucleons in a $2N$ ($3N$) system can be labelled by an index $i\!=\!1,2,(3)$. A single-nucleon operator with subindex $i$ is understood to act on nucleon $i$. 
The leading single-nucleon contributions to $J^{\mu}$ and  the leading single-nucleon contributions induced by the terms 
in  the first line of Eq.\,(\ref{eq:impcoup}) to $J_{\CPv}^{\mu}$ 
are~\footnote{Here and in the following the elementary charge $e$ is defined to be negative, $e<0$.}
\begin{equation}\label{eq:ncurrent}
 J^{\mu}_i=-\frac{e}{2}\left(1+\tau^3_{(i)} \right)\, v^{\mu}\,\,,\qquad 
 J^{\mu}_{\CPv,i}=-\frac{1}{2}\left[d_n \left(1-\tau^3_{(i)}\right)+d_p\left(1+\tau^3_{(i)}\right)\right]\,i\vec{q}\cdot \vec{\sigma}_{(i)}\, v^{\mu}\,\,.
\end{equation}
Other irreducible CP-conserving and CP-violating current operators only contribute to $\tilde{J}^{\mu}$ at \NtwoLO\ as 
discussed in~\cite{deVries2011b, Bsaisou:2012rg,dissertation} and are thus irrelevant for this work. 

\begin{figure}[t]
\centering
\includegraphics[scale=0.6]{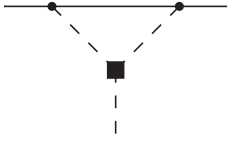}
\caption{Correction to the $g_1$ pion-nucleon vertex induced by the CP-violating three-pion vertex in Eq.\,(\ref{eq:impcoup}). Notation as in Fig.~\ref{Fig:current}. \label{Fig:g1corr}}
\end{figure}

For $2N$ operators we use the definitions
 $\vec{\sigma}_{(ij)}^{\pm}:=\vec{\sigma}_{(i)}\pm\vec{\sigma}_{(j)}$, $\tau_{(ij)}^{\pm}:=\tau^3_{(i)}\pm\tau^3_{(j)}$ 
 with $i\!\neq\! j$ and $\vec{k}_{i}:=\vec{p}_i-\vec{p}_i^{\,\prime} $, where  $\vec{p}_i\,(\vec{p}_i^{\,\prime}$) 
 is the momentum of an incoming (outgoing) nucleon. 
 The leading $2N$ irreducible potential operators induced by the terms 
 in Eq.\,(\ref{eq:impcoup}) are \cite{khriplovich,Afnan,Bsaisou:2012rg,Maekawa:2011vs}
\begin{eqnarray}\label{eq:nnpot}
  V_{\CPv,ij}^{NN}(\vec{k}_{i}) = 
   &&i\frac{g_A}{2\fpi}\frac{\vec{k}_{i}}{\vec{k}_{i}^{\,2}+\mpi^2}\ g_0\,\vec{\sigma}_{(ij)}^{-}\vec{\tau}_{(i)}\cdot\vec{\tau}_{(j)}\nn \\
&+&i\frac{g_A}{4\fpi}\frac{\vec{k}_{i}}{\vec{k}_{i}^{\,2}+\mpi^2}\,
\left[g_1+\Delta\, \,f_{g_1}(|\vec k_i|) \right] \left(\vec{\sigma}_{(ij)}^{+}\tau_{(ij)}^{-}+\vec{\sigma}_{(ij)}^{-}\tau_{(ij)}^{+} \right)\nn \\
&+&\frac{i}{2}\frac{\beta^2\mpi^2\vec{k}_{i}}{\vec{k}_{i}^{\,2}+\beta^2\mpi^2}\,
\Bigl [C_1\vec{\sigma}_{(ij)}^{-}+C_2\,\vec{\sigma}_{(ij)}^{-}\vec{\tau}_{(i)}\cdot\vec{\tau}_{(j)}\Bigr]\,.
\end{eqnarray}
Here $g_A= 1.269$ is the axial-vector coupling constant of the nucleon, 
$\fpi = (92.2\pm0.1)\,\rm{MeV}$  the pion decay constant and $\mpi=138.01$ MeV the 
isospin-averaged pion mass \cite{pdg}. 
When presenting numerical results, the limit $\beta\to \infty$ is chosen. 
The parameter $\beta$ is introduced only as a diagnostic tool to compare our $\chi$EFT results with those based on phenomenological potentials.
The $g_1$ vertex correction $\Delta\,f_{g_1}$ is induced by the 
three-pion $\Delta$ vertex in Eq.\,(\ref{eq:impcoup}) via the finite one-loop diagram depicted in Fig.~\ref{Fig:g1corr}. 
This diagram yields \cite{deVries:2012ab}
\begin{equation}\label{eq:g1corr}
f_{g_1}(k) \equiv \, -\frac{15}{32}\frac{g_A^2\mpi \mn}{\pi\fpi^2}\left[1+ \left(\frac{1+2\vec{k}^{\,2}/(4\mpi^2)}{3 |\vec{k}\,|/(2\mpi)}\arctan\left(\frac{|\vec{k}\,|}{2\mpi}\right)-\frac{1}{3}\right)\right]\,,
\end{equation}
where the terms within the brackets have been arranged to indicate the constant and $k$-dependent components, respectively. The dominant $k$-independent component of $f_{g_1}$ is larger by a factor of $5 \pi$, roughly an order of magnitude, than the power-counting estimate. The enhancement by a numerical factor of $\pi$ is a common feature of triangular diagrams \cite{Bernard:1991rt,Becher:1999he,Friar:2003yv,Liebig:2010ki,Baru:2012iv}, while the factor~5 can be traced back to a coherent sum over isospin.

The three-pion $\Delta$ vertex also gives  rise to the leading irreducible CP-violating $3N$ potential relevant for the considered $3N$ systems \cite{deVries:2012ab,dissertation},
\begin{eqnarray}\label{eq:3npot}
V_{\CPv}^{3N}(\vec{k}_1,\vec{k}_2,\vec{k}_3)
&=&-i \Delta \frac{\mn g_A^3}{4\fpi^3}
 \left(\delta^{ab}\delta^{c3}+\delta^{ac}\delta^{b3}+\delta^{bc}\delta^{a3}\right)\tau_{(1)}^{a}\tau_{(2)}^{b}\tau_{(3)}^{c}\nn \\
&&\mbox{}\times\frac{ (\vec{\sigma}_{(1)}\cdot\vec{k}_{1})
                                    (\vec{\sigma}_{(2)}\cdot\vec{k}_{2})
                                    (\vec{\sigma}_{(3)}\cdot\vec{k}_{3})}
                                   {\left[\vec{k}_1^{\,2}+\mpi^2\right]
                                    \left[\vec{k}_2^{\,2}+\mpi^2\right]
                                    \left[\vec{k}_3^{\,2}+\mpi^2\right]}\,\,,
\end{eqnarray}
where $a,b,c$ are isospin indices and $\delta^{ab}$ is the Kronecker delta.
The full CP-violating potential operator $V_{\CPv}$ is in general the sum of $V_{\CPv,ij}^{NN}$, for all permutations of nucleon indices, and $V_{\CPv}^{3N}$.

\setcounter{equation}{0}
\section{The EDMs of the deuteron, the helion and the triton}
\label{sec:nucedms}
The CP-conserving wave functions of the deuteron have been computed by solving the scattering equations 
with the \NtwoLO\ chiral potential  of Refs. \cite{Epelbaum:2004fk,chiralpotentials} for the following five combinations of Lippmann-Schwinger cutoffs $\Lambda_{{\rm LS}}$ and Spectral-Function-Regularization cutoffs $\Lambda_{{\rm SFR}}$ (see Refs.~\cite{Epelbaum:2004fk,chiralpotentials} for a detailed explanation of these cutoffs):
\begin{equation}\label{eq:cutoffs}
 (\Lambda_{{\rm LS}},\Lambda_{{\rm SFR}})=\Bigl\{ (0.45,0.5);  (0.6,0.5); 
 (0.55,0.6); (0.45,0.7);
 (0.6,0.7) \Bigr\}\, {\rm GeV}\,\,.
 \end{equation}
 
The CP-conserving interactions are here treated non-perturbatively in all orders. Such an approach requires necessarily 
that cutoffs can only be varied in a limited range. We note that a perturbative treatment of higher orders removes this constraint 
\cite{Valderrama:2011hz,Valderrama:2011hw}. The results are, however, equivalent as long as the cutoffs are of the order of 0.5~GeV,
which is completely sufficient for this work. With the used range of cutoffs, several LECs of the  \NtwoLO\ chiral potential 
change sign. Therefore, we are confident that the employed range of cutoffs is suitable for reliable uncertainty estimates of the 
CP-violating contributions.
 
In order to compare our results with previous ones computed from {\em phenomenological} CP-conserving  potentials, the A$v_{18}$ 
 (A$v_{18}$+UIX) potential \cite{av18,Pudliner:1997ck} and the CD-Bonn (CD-Bonn+TM) potential \cite{cdbonn,Coon:2001pv} have also been applied for the deuteron ($3N$ cases).

The single-nucleon contributions to the deuteron EDM are given by the sum of the neutron and proton 
EDMs \cite{LiuTimmermans}
as indicated in the first two rows of Table~\ref{tab:deuteron}.
The deuteron wave function has a ${}^3\!S_1$ and a small ${}^3\!D_1$ component and its isospin is $I\!=\!0$. Since the leading contribution to $J^{\mu}$ 
(see Eq.\,(\ref{eq:ncurrent})) is spin independent, the convolution of the deuteron wave function 
with $V_{\CPv}$ 
of Eq.\,(\ref{eq:nnpot}) has to yield a ${}^3\!P_1$ intermediate state with $I\!=\!1$ in order 
to obtain a nonvanishing complete nuclear matrix element of $\tilde{J}^{\mu}$~\cite{LiuTimmermans}. 
Only the terms proportional to $g_1$ and $\Delta f_{g_1}$ in Eq.\,(\ref{eq:nnpot}) fulfill this isospin selection rule. 
Their contributions are given in the last two rows of Table~\ref{tab:deuteron}. 
The momentum-independent component of $f_{g_1}(\vec{k})$, 
as defined by the first term in the brackets of Eq.\,(\ref{eq:g1corr}), 
amounts to approximately 90\% of the total contribution from the $\Delta$-induced $g_1$ vertex correction.

The listed EDM contributions of the $\chi$EFT potentials are given by the center  of the interval resulting
from the different cutoff combinations. The pertinent uncertainty is determined from  the difference between the center and the boundaries of the interval. We will call this type of uncertainty the {\em nuclear} uncertainty in order to distinguish it from the {\em hadronic} uncertainty
which is related to the  {\em low-energy} coefficients appearing in Eq.~(\ref{eq:impcoup}). The results for the phenomenological potentials considered
are also shown in Table~\ref{tab:deuteron} and agree (where a comparison is possible) with those in 
Refs.~\cite{khriplovich,Pospelov_deuteron,LiuTimmermans,Stetcu:
2008vt,Afnan,deVries2011a,deVries2011b,Bsaisou:2012rg,Song:2012yh}. 
The values from chiral and phenomenological potentials are in excellent agreement. 

\renewcommand{\thefootnote}{\fnsymbol{footnote}}
\begin{table*}[t]
\begin{center}
\caption[Contributions to  the deuteron EDM]{Contributions 
to  the deuteron EDM from the \NtwoLO\ $\chi$EFT 
potential~\cite{Epelbaum:2004fk,chiralpotentials}, the A$v_{18}$  potential \cite{av18} and the CD-Bonn  potential \cite{cdbonn}, respectively. The results from the \NtwoLO\ $\chi$EFT potential are defined as the center of each interval obtained by employing the five different combinations of LS and SFR cutoffs given in Eq.\,(\ref{eq:cutoffs}). The corresponding {\em nuclear} uncertainty is  the difference between this central value and the  boundary values. 
The results still have to be multiplied by the corresponding coefficients in Eq.~(\ref{eq:impcoup}) which here are included in the units.
Note that $d_n$, $d_p$ carry themselves dimensions, namely $[e\,\rm{fm}]$, while  $g_1$  and $\Delta$ are dimensionless.\footnotemark[4]\
\label{tab:deuteron}}

\smallskip
\begin{tabular}{c || c | c | c || c}
label & \NtwoLO\ $\chi$EFT& A$v_{18}$  & CD-Bonn & units \\
\hline
\hline
$d_{n}$               & $1.00$ & $\phantom{-}1.00$ & $\phantom{-}1.00$ & $d_n$\\
$d_{p}$               & $1.00$ & $\phantom{-}1.00$ & $\phantom{-}1.00$ & $d_p$\\
$g_1$                  & $ -0.183\pm 0.017 $                 & $-0.186$                  & $-0.186$ & $g_1\,e\,{\rm fm}$ \\
$\Delta\,f_{g_1}$ & $\phantom{-}0.748\pm0.138$   &$\phantom{-}0.703$ &$\phantom{-}0.719$ & $\Delta \,e\,{\rm fm}$
\end{tabular}
\end{center}
\end{table*}
\footnotetext[4]{For the $D$-wave corrected
form of the weights of $d_n$ and $d_p$  in Table~\ref{tab:deuteron} see Appendix~\ref{app:erratum}.}
\renewcommand{\thefootnote}{\arabic{footnote}}

The wave functions of the helion and triton have been computed by solving the Faddeev equations for the considered CP-conserving potentials. By a series of arithmetic manipulations~\cite{dissertation}, the second and third term on the right-hand side of Eq.\,(\ref{eq:ptvcurrent}) lead to Faddeev equations which have also been solved numerically. Within this computation both $I=1/2$ and $I=3/2$ components of the helion and triton wave functions (with total angular momentum $J=1/2$) as well as electromagnetic interactions have been considered. 
The strict isospin selection rule of the deuteron is absent in the helion and triton cases due to a significantly larger number of wave function components and possible intermediate states. All operators in Eq.\,(\ref{eq:nnpot}) and Eq.\,(\ref{eq:3npot}) yield non-vanishing EDM contributions for these nuclei.
%
\begin{table*}[tbh!]
\begin{center}
\caption{Contributions to the helion and triton EDMs from the \NtwoLO\ $\chi$EFT potential with three-nucleon forces \cite{Epelbaum:2004fk,chiralpotentials}, the A$v_{18}$+UIX potential \cite{av18,Pudliner:1997ck} and the CD-Bonn+TM potential \cite{cdbonn,Coon:2001pv}, respectively. The results are presented as
in Table~\ref{tab:deuteron}. 
Note that $d_{n,p}$ carry dimension $[e\,\rm{fm}]$ and $C_{i}$ dimension $[{\rm fm}^3]$.
\label{tab:3n}}

\smallskip
\begin{tabular}{c|| c || c | c | c || c}
          label    & A &\NtwoLO\ $\chi$EFT& A$v_{18}$+UIX & CD-Bonn+TM & units\\\hline
\hline
$d_n$    & $\he$ &$\phantom{-}0.904\pm0.013$&$\phantom{-}0.875$&$\phantom{-}0.902$& $d_n$\\
              & $\hy$&$-0.030\pm0.007$&$-0.051$&$-0.038$& $d_n$\\
\hline
$d_p$    & $\he$ &$-0.029\pm0.006$&$-0.050$&$-0.037$& $d_p$\\
              & $\hy$&$\phantom{-}0.918\pm0.013$&$\phantom{-}0.902$&$\phantom{-}0.876$& $d_p$\\
\hline
$\Delta $ & $\he$&$\phantom{-}0.017\pm0.006$&$\phantom{-}0.015$&$\phantom{-}0.019$& $\Delta\, e\, {\rm fm}$ \\
              & $\hy$&$\phantom{-}0.017\pm0.006$&$\phantom{-}0.015$&$\phantom{-}0.018$& $\Delta\,  e\, {\rm fm}$\\
\hline
$g_0$    & $\he$&$-0.111\pm0.013$&$-0.073$&$-0.087$&$ g_0\,e\,{\rm fm}$\\
              & $\hy$&$\phantom{-}0.108\pm0.013$&$\phantom{-}0.073$&$\phantom{-}0.085$&$ g_0\,e\,{\rm fm}$\\
\hline
$g_1$    & $\he$&$-0.142\pm0.019$&$-0.142$&$-0.146$&$ g_1\,  e\,{\rm fm}$\\
              & $\hy$&$-0.139\pm0.019$&$-0.142$&$-0.144$&$ g_1\,  e\,{\rm fm}$\\
\hline
              $\Delta\, f_{g_1}$   & $\he$ &$\ \ 0.608\pm0.142$&$\ \ 0.556$&$\ \ 0.586$&$ \Delta\, e\,{\rm fm}$\\
              & $\hy$&$\ \ 0.598\pm0.141$&$\ \ 0.564$&$\ \ 0.576$&$ \Delta \,e\,{\rm fm}$\\              
\hline
$C_1$   & $\he$&$\phantom{-}0.042\pm0.017$&$\phantom{-}0.0014$&$\phantom{-}0.016$&$ C_1 \,e\,{\rm fm}^{-2}$\\
              & $\hy$&$-0.041\pm0.016$&$-0.0014$&$-0.016$&$C_1\, e\,{\rm fm}^{-2}$\\
\hline              
$C_2$   & $\he$ &$-0.089\pm0.022$&$-0.0042$&$-0.033$&$ C_2\, e\,{\rm fm}^{-2}$\\
              & $\hy$&$\phantom{-}0.087\pm0.022$&$\phantom{-}0.0044$&$\phantom{-}0.032$&$ C_2 \,e\,{\rm fm}^{-2}$\\
\end{tabular}
\end{center}
\end{table*}

The EDM results are listed in Table \ref{tab:3n} for all  CP-conserving potentials considered. 
For the phenomenological potentials, the EDM contributions induced by $d_{n,p}$ and $g_{0,1}$ are in agreement with those of Ref.~\cite{Song:2012yh}, while the $g_{0,1}$-induced contributions are smaller than those of Refs.~\cite{deVries2011b,Stetcu:2008vt} by a factor of two. The dependence of the contributions induced by the $C_i$ vertices on the cutoff parameter $\beta$ defined in Eq.\,(\ref{eq:nnpot}) is discussed in Appendix~\ref{app:beta} explicitly for the case of $C_2$. The contributions induced by the short-range $C_i$ vertices when the A$v_{18}$+UIX potential is employed are smaller than the corresponding results reported in Ref.\cite{deVries2011b}. This discrepancy might partially be attributed to a deviation similar to the one mentioned before for the $g_{0,1}$-induced contributions. The slow convergence found in Ref.\cite{deVries2011b} for $\chi$EFT potentials could not be confirmed within our approach for the \NtwoLO\ $\chi$EFT potential utilized here. The $d_{n,p}$ and $g_{0,1}$ contributions from the chiral potential and the phenomenological potentials are in reasonable agreement. The largest difference occurs for the $g_0$ contribution with a $20\%-30\%$ enhancement for the chiral potential.

Next we discuss the contributions from the three-pion $\Delta$ vertex which is considered for the first time in this paper. In contrast to the power-counting estimate, the by far dominant contributions arise from the loop-induced $g_1$ vertex correction. These contributions are larger by roughly a factor of $50$ than the contributions from the three-body potential in Eq.~\eqref{eq:3npot}. This discrepancy can only partially be attributed to the enhancement of the one-loop diagram by the factor of $5\pi$ mentioned before. Furthermore, the power-counting estimates of the $g_{0,1}$-induced potential operators modulo $g_{0,1}$ equal the one of the $3N$ CP-violating potential modulo $\Delta$ according to 
Appendix~\ref{app:pc}. The explicit computation of the three-body $\Delta$ term yields a contribution that is approximately one order of magnitude smaller than our power-counting scheme predicts. 
We investigated whether this suppression is related to the high symmetry of the $3N$ wave functions, which are usually dominated by a completely antisymmetric spin-isospin state. However, taking only this principal $S$-state into account, we found
$(0.016\pm0.04)\,\Delta\,e\,{\rm fm}$ for the \NtwoLO\ $\chi$EFT potential,  $0.016\,\Delta\,e\,{\rm fm}$ for A$v_{18}$+UIX potential and $0.018\,\Delta\,e\,{\rm fm}$  for CD-Bonn+TM potential, which are very similar to the full results.
Therefore, the origin of the discrepancy to the power-counting estimate is not known to us at this point.

The contributions from the short-range $C_i$ vertices, $i=1,2$, are highly model-dependent as the last four rows 
of Table~\ref{tab:3n} indicate. The A$v_{18}$(+UIX) and CD-Bonn(+TM) results differ by one order in magnitude and are themselves smaller than the results obtained by employing the \NtwoLO\ $\chi$EFT potential. 
The small value of the $C_{i}$--contributions for the A$v_{18}$(+UIX) cases results from an atypically pronounced short-range repulsion of the A$v_{18}$(+UIX) potential. This can be verified by evaluating the pertinent diagrams with a finite value of $\beta$, where $\beta$ denotes the mass
scale in a form factor that is attached to the four--nucleon vertex, {\it cf.} Eq.\,(\ref{eq:nnpot}). For $\beta\simeq 3$, which corresponds to cutoffs of the order of $0.5$ GeV, the A$v_{18}$(+UIX) results are in line with the \NtwoLO\ $\chi$EFT ones. Such cutoffs are standard for implementations of chiral interactions because they lead to natural-sized LECs for four-nucleon vertices. The addition of such a cutoff does not significantly alter the $\chi$EFT results  --- see also Appendix~\ref{app:beta}.

\renewcommand{\thefootnote}{\fnsymbol{footnote}}
On the basis of  the four, respectively, eight  chiral results of Table~\ref{tab:deuteron} and Table~\ref{tab:3n}, we find  the following predictions for the deuteron, helion and triton EDMs which depend on the  low-energy constants of the Lagrangian given in Eq.\,(\ref{eq:impcoup}):\footnotemark[4]
\footnotetext[4]{For the $D$-wave corrected
form  of the $d_n$ and $d_p$ weights in  Eq.\,(\ref{eq:h2edm}) see Appendix~\ref{app:erratum}.}
\renewcommand{\thefootnote}{\arabic{footnote}}
\begin{eqnarray}
d_{^2{\rm H}} &=&
d_n + d_p 
- \bigl [ (0.183 \pm 0.017) \,g_1
- (0.748\pm 0.138) \,\Delta\bigr] \,e \,{\rm fm} \,\,,
 \label{eq:h2edm} 
 \\
d_{{}^3{\rm He}} &=& (0.90 \pm 0.01)\, d_n-(0.03\pm 0.01 )\,d_p \nn \\
&&\mbox{}+
\bigl\{ (0.017\pm0.006)\,\Delta - (0.11\pm 0.01)\,g_0  -(0.14\pm 0.02)\,g_1  +(0.61\pm 0.14)\,\Delta\nn\\
&&\quad\mbox{}+ \left[ (0.04\pm 0.02)C_1 -(0.09\pm 0.02)C_2 \right]\times  {\rm fm}^{-3} \bigr\}
\, e\, {\rm fm} \,\, , \label{eq:he3edm}
\\
d_{{}^3{\rm H}}&=& -(0.03\pm0.01)\, d_n + (0.92\pm 0.01)\,d_p \nn \\
&&\mbox{}+
\bigl\{  (0.017\pm0.006)\,\Delta + (0.11\pm 0.01)\,g_0  - (0.14\pm 0.02)\,g_1\mbox{} +(0.60\pm 0.14)\,\Delta \nn\\
&&\quad\mbox{}- \left[(0.04 \pm 0.02) C_1 -(0.09\pm 0.02) C_2\right] \times  {\rm fm}^{-3} \bigr\}
\, e\, {\rm fm}\,\, . \label{eq:h3edm}
\end{eqnarray}
The numbers presented here do not in all cases agree with the power-counting estimates in Appendix~\ref{app:pc}. The one-pion-exchange contributions in particular are smaller than the power counting predicts by roughly a factor $3$--$5$, which was also found in Ref.~\cite{deVries2011b}. The main consequence is that the short-range contributions proportional to $C_{1,2}$, which are roughly in agreement with their power-counting estimates, become relatively more important. As discussed above, the three-pion-exchange contributions proportional to $\Delta$ are also smaller than expected. We do not know the reason for these discrepancies to the power counting which has been otherwise so successful in many CP-conserving processes.
Although the numbers in  the Tables~\ref{tab:deuteron} and \ref{tab:3n}
are not always in line with the power-counting estimates, explicit calculations~\cite{deVries2011b,Bsaisou:2012rg,dissertation} revealed that subleading
corrections are  indeed suppressed compared with the results
listed in 
Eqs.~(\ref{eq:h2edm})--(\ref{eq:h3edm}). The nuclear uncertainties of the latter terms may be reduced by the replacement of the
\NtwoLO\ CP-conserving chiral
potentials and pertinent wave functions by their \NthreeLO\ counter parts.

The above results for the deuteron, helion, and triton EDMs hold regardless of the underlying mechanism of CP violation. In order to continue the analysis, a particular source ({\it e.g.} the $\theta$-term, a quark (chromo-)EDM etc.) or a CP-violating high-energy model (in Ref.~\cite{Dekens:2014jka} the minimal left-right symmetric model (mLRSM) and the aligned two-Higgs doublet model were studied) has to be specified. The coefficients of Eq.\,(\ref{eq:impcoup}) can then be 
calculated -- in the future by Lattice QCD --  or estimated within such a particular scenario in order to identify the hierarchies of contributions to the various EDMs, see {\it e.g.} the analysis in Ref.~\cite{Yamanaka:2014nba}. In order to focus on the cases of the QCD $\theta$-term and the mLRSM in the two subsequent sections, the discussion 
in Ref.~\cite{Dekens:2014jka} is briefly repeated and updated.

\setcounter{equation}{0}
\section{EDMs of light nuclei from the QCD \boldmath{$\theta$}-term}
\label{sec:theta}
\subsection{Estimates of the coupling constants in the \boldmath{$\theta$}-term scenario}
\label{sec:theta_coeff}
The QCD $\theta$-term can be removed by an axial $U(1)$ transformation at the price of picking up a complex phase of the quark-mass matrix \cite{BiraEmanuele}:
\begin{equation}
\mathcal{L}_{{\rm QCD}}=\cdots+
\frac{m_um_d}{m_u+m_d}\,\bar{\theta}\,\bar{q}i\gamma_5 q\,\,.
\end{equation}
Since the leading low-energy constants (LECs) of chiral perturbation theory and its heavy-baryon extensions induced by the quark-mass matrix are quantitatively known, the coefficients $\Delta$, $g_0$, and $g_1$ in Eq.\,(\ref{eq:impcoup}) can be related to quantitatively known matrix elements in the $\theta$-term case \cite{BiraEmanuele,Bsaisou:2012rg,dissertation}.
In particular, the three-pion $\Delta$ vertex can be related to the strong part of the pion-mass splitting \cite{BiraEmanuele,dissertation}, 
\begin{equation}\label{eq:Delta_theta}
 \Delta^\theta=\frac{ \epsilon (1-\epsilon^2)}{16 \fpi \mn} \frac{M_\pi^4}{M_K^2 - M_\pi^2} \,\bar\theta +\cdots= ( -0.37\pm 0.09) \cdot 10^{-3} \, \bar\theta\,,
 \end{equation}
with the average kaon mass $M_K=494.98$ MeV \cite{pdg} and $\epsilon \equiv (m_u-m_d)/(m_u+m_d)=-0.37\pm0.03$ computed from the latest prediction of $m_u/m_d=0.46\pm0.03$ of 
Ref.~\cite{Aoki:2013ldr}. 
The dots in Eq.~(\ref{eq:Delta_theta}) 
denote higher-order contributions which are included in the uncertainty estimate.
The isospin-breaking pion-nucleon coupling constant $g_1$ has two leading contributions. The first arises from a shift of the ground state due to the $\theta$-term and is given by
\begin{equation}\label{eq:g1_theta}
g_1^\theta(c_1) = 8 c_1 \mn \Delta^\theta = (2.8 \pm 1.1)\cdot 10^{-3} \, \bar\theta\,\,,
\end{equation}
which corresponds to $(-7.5\pm2.3)\Delta^\theta$ and where 
$c_1=(1.0 \pm 0.3)\,{\rm GeV}^{-1}$~\cite{Baru:2011bw} is related to the nucleon sigma term.
For details we refer to Refs.~\cite{BiraEmanuele,Bsaisou:2012rg}. 
The second contribution, labelled $\tilde g_{1}^\theta$, is currently not quantitatively assessable. It was estimated in Ref. \cite{Bsaisou:2012rg} 
by resonance saturation to equal
 $\tilde g_{1}^\theta= (0.6 \pm 0.3) \cdot 10^{-3}\cdot \bar\theta $, while its Naive Dimensional Analysis (NDA) estimate, {\it cf.} Ref.~\cite{BiraEmanuele}, is 
 $|\tilde g_1^\theta| \sim  |\epsilon| {M_\pi^4}/{(\mn^3 \fpi)} \sim 1.7 \cdot 10^{-3} \cdot
 \bar\theta$.  These estimates can be combined by regarding the result from resonance saturation as the central value and the difference to the NDA estimate as the uncertainty, which yields
 \begin{equation}\label{eq:tilde_g1_theta}
 \tilde g_1^\theta = (0.6 \pm 1.1)\cdot 10^{-3}\, \bar\theta\,\,.
 \end{equation}
 This contribution has to be added to $g_1^\theta(c_1)$ to obtain the total value of the $g_1^\theta$ coupling constant:
 \begin{equation} \label{eq:g1_theta_solo}
  g_1^\theta  = g_1^\theta(c_1) + \tilde g_1^\theta = (3.4 \pm 1.5)  \cdot 10^{-3}\, \bar\theta\,\,.
 \end{equation}
 
The coefficient of the isospin-conserving CP-violating $\pi N$ vertex, $g_0^\theta$, is interrelated with the quantitatively known strong contribution to the neutron-proton mass shift,
$\delta m_{np}^{\rm str}$. We do not apply the value for $\delta m_{np}^{\rm str}$ used in  Refs.~\cite{Bsaisou:2012rg,dissertation,Dekens:2014jka} here, but instead the more refined value $\delta m_{np}^{\rm str} = (2.44\pm 0.18)\,{\rm MeV}$, which follows from a weighted
average of the values compiled in Ref.\,\cite{Walker-Loud:2014iea} and the newest lattice result of 
Ref.~\cite{Borsanyi:2014jba}. We then obtain
\begin{equation} \label{eq:g0_theta}
  g_0^\theta  =\bar\theta 
  \frac{\delta m_{np}^{\rm str} (1- \epsilon^2)}{4 \fpi \epsilon}
  = (-15.5 \pm 1.9) \cdot 10^{-3} \, \bar\theta\,,
\end{equation}
where the latest update of Ref.~\cite{Aoki:2013ldr} for the value of $\epsilon$, see above, has also been included.

The neutron and proton EDMs induced by the $\theta$-term have recently been calculated in 
Refs.~\cite{Guo12,Akan:2014yha} on the basis of supplementary Lattice-QCD 
input~\cite{Shintani:2008nt,Shintani:2012zca,Shintani:2014},
\begin{equation}
d_n^{\theta}= \bar\theta\cdot(2.7\pm1.2)\cdot 10^{-16}\, e\,{\rm cm}\,\, ,\qquad d_p^{\theta}=-\bar\theta\cdot(2.1\pm 1.2)\cdot 10^{-16}\, e\,{\rm cm} \,\, ,
 \label{eq:edm_n_theta}
\end{equation}
where the signs have been adjusted to our convention $e<0$.

The coefficients of the nucleon-nucleon contact interactions, $C_{1,2}$, are harder to quantify.
In principle they could be deduced from an analysis of isospin-violating pion production in $NN$ collisions studied in Ref.~\cite{Filin:2009yh}
since the CP--violating nucleon-nucleon contact terms are related to the isospin--violating $NN\to NN\pi$ contact terms in very much the same way
as is the $g_0$--term to the proton--neutron mass difference.\footnote{This would require a calculation of the isospin-violating $NN\to NN\pi$ amplitudes to one order higher than currently performed as well as an additional analysis of
the reaction $dd\to \alpha\pi^0$  --  the status of the theory for this reaction is reported in Ref.~\cite{Baru:2013zpa} and the latest data
is presented in Ref.~\cite{Adlarson:2014yla}. } 
However, this analysis has not yet been performed with the necessary accuracy.
We therefore estimate the strengths of the $C_{1,2}$ via  the power-counting estimates of the $g_0$--induced two-pion-exchange diagrams since these coefficients should absorb the divergences and associated scale dependences of such diagrams. This procedure yields the estimate 
\begin{equation}\label{eq:C_i}
|C_{1,2}^{\theta}|=\mathcal{O}\left(\gzero g_A /(\fpi \mn^2)\right) \simeq 2\, \bar{\theta}\cdot 10^{-3} \, {\rm fm}^3\,\,, 
\end{equation}
while the signs of $C_{1,2}^{\theta}$ remain unknown. Therefore, the $C_{1,2}$-induced contributions and their nuclear 
uncertainties 
will be added in quadrature to provide an additional -- and difficult to reduce --  uncertainty to the total EDM results.

\subsection{Results for the deuteron and \boldmath{$3N$} EDMs in the \boldmath{$\theta$}-term scenario} 

We can now insert the above predictions for the coefficients in Eq.\,(\ref{eq:impcoup}) into the power-counting estimates presented in Appendix~\ref{app:pc}. The following hierarchy of nuclear EDM contributions then emerges for the deuteron case:
The $\gzero$-induced one-pion exchange and the $C_{1,2}$-induced contact interactions vanish in the $2N$ system of the deuteron due to isospin selection rules. Therefore, the leading-order EDM contribution is defined by the $\gone$-induced one-pion exchange \cite{deVries2011a,Bsaisou:2012rg}. The $\gone$ vertex is corrected by the $\Delta$-dependent term in Eq.\,(\ref{eq:nnpot}). It generates EDM contributions which are approximately half the size of the tree-level ones induced by the $\gone$ vertex but with the same sign. The only other relevant EDM contribution up to \NLO\ ({\it i.e.} contributions suppressed by a factor of $\sim\mpi/\mn$) is the 
isoscalar sum of the single-nucleon EDMs~\cite{Ottnad:2009jw,Mer11,deVries2011b,Bsaisou:2012rg}. In this combination, the large isovector loop contribution to the single-nucleon 
EDM cancels~\cite{CDVW79}. 

\renewcommand{\thefootnote}{\fnsymbol{footnote}}
Thus the EDM of the deuteron generated by the $\theta$-term up-to-and-including \NLO\ is given 
by the insertion of $d_n^{\theta}$, $d_p^{\theta}$, $\gone$ and $\Delta^{\theta}$ 
into Eq.\,(\ref{eq:h2edm}):\footnotemark[4]\footnotetext[4]{For the $D$-wave corrected
form  of the $d_n$ and $d_p$ weights contributing to  Eq.\,(\ref{eq:d2Htheta}) and appearing in
 Eq.\,(\ref{dEDMtheta}) see Appendix~\ref{app:erratum}.}
\begin{equation}
d^\theta_{{}^2{\rm H}} =
   \bar\theta\cdot \left\{  \bigl[(0.6 \pm 1.7)\bigr] 
  - (0.62 \pm 0.06\pm 0.28) - (0.28\pm 0.05\pm 0.07)
   \right\}\cdot 10^{-16}\,  e\, {\rm cm}\,\,.
    \label{eq:d2Htheta}
\end{equation}
In each set of parentheses, the first (second) 
uncertainty is the nuclear (hadronic) one,
except for the sum of single-nucleon contributions, which here
only have hadronic uncertainties. 
Because of the rather large uncertainty of the sum of single-nucleon contributions, the pure two-body contribution\footnotemark[4]\renewcommand{\thefootnote}{\arabic{footnote}}
\begin{equation}\label{dEDMtheta}
d^\theta_{{}^2{\rm H}} - d_p^\theta -d_n^\theta =- \bar\theta\cdot (0.89 \pm 0.30) \cdot 10^{-16}\,  e\, {\rm cm}\,\,,
\end{equation}
where 
the 
uncertainties have been added in quadrature, is more useful to consider. This expression can be applied to extract $\bar\theta$ from the measurements of the proton, neutron and deuteron EDMs without
additional theoretical input. The contribution of the $\Delta$-induced $g_1$ vertex correction was not considered in Ref.~\cite{Dekens:2014jka}, where therefore a $50\%$ smaller result for the total two-body contribution was obtained.

For the $3N$ systems, the \LO\ EDM contributions -- apart from the single-nucleon ones   --
are defined by the $\gzero$-induced one-pion exchange as depicted in Fig.~\ref{Fig:current}~(b). The $\gone$-induced one-pion exchange is counted as 
\NLO\ since $g_1^\theta/g_0^\theta = - 0.22 \pm 0.10$. The $\Delta$-dependent correction to $g_1$ yields contributions which are, as in the deuteron case, roughly one half of the $\gone$ contributions with the same sign. As discussed before, the $3N$ contributions proportional to $\Delta$ are smaller than estimated by power counting and 
are negligible, while
the contributions from the CP-violating nucleon-nucleon vertices are accounted for as additional overall uncertainties.

In the order of the rows of Table \ref{tab:3n} and the terms in Eqs.\,(\ref{eq:he3edm})-(\ref{eq:h3edm}), the different contributions to the helion and triton EDMs induced by the $\theta$-term, respectively,  combine to the following total helion and triton EDMs:
\begin{eqnarray}
d_{{}^3{\rm He}}^{\theta}&=&
\bar\theta\cdot \bigl\{  \left [(2.44 \pm 0.04 \pm 1.08) + (0.06 \pm 0.01 \pm  0.03) \right] \nn \\
&&\quad\mbox{}
 -(0.006\pm 0.002 \pm 0.001)+  (1.72\pm 0.20 \pm 0.21)
 -(0.48 \pm 0.06\pm 0.22) \nn \\
&&\quad\mbox{} -(0.22\pm 0.05\pm 0.05) \pm 0.2 \bigr\}
\cdot 10^{-16}\, e\,{\rm cm} \nn \\
&=&\bar\theta\cdot( 3.5 \pm 1.2)\cdot 10^{-16}\,e\, {\rm cm}\,\,, \\[2mm]
d_{{}^3{\rm H}}^{\theta}&=&
\bar\theta\cdot \bigl\{  \left [ - (0.08 \pm 0.02 \pm  0.04)-(1.93 \pm 0.03 \pm 1.10) \right] \nn \\
&&\quad\mbox{}
-(0.006\pm0.002\pm0.001)-  (1.68\pm 0.20 \pm 0.21)
 -(0.47 \pm 0.06\pm 0.21) \nn \\
&&\quad\mbox{} -( 0.22\pm 0.05\pm 0.05) \pm 0.2 \bigr\}
\cdot10^{-16}\,e\, {\rm cm} \nn \\
&=&-\bar\theta\cdot(4.4 \pm 1.2)\cdot10^{-16}\,e\, {\rm cm}\,\, .
\end{eqnarray}
In each set of parentheses, the first uncertainty is always the nuclear one, while the second is the hadronic one.
In order to remove the influence of the single-nucleon EDM values which rely on Lattice-QCD input at still rather large quark masses, we also list the pure multi-body contributions to the EDMs
where the nuclear uncertainty of the single-nucleon terms can safely be neglected:
\begin{eqnarray}\label{triEDMtheta}
d_{{}^3{\rm He}}^{\theta}-0.90\, d_n^{\theta}+0.03 \,d_p^{\theta}
&=&\phantom{-} \bar\theta\cdot (1.01\pm 0.42)\cdot 10^{-16}\, e\, {\rm cm}\,\, ,\nonumber\\
d_{{}^3{\rm H}}^{\theta} -0.92\,d_p^{\theta} +0.03\,d_n^{\theta}
&=&-\bar\theta\cdot (2.37 \pm 0.42)\cdot10^{-16}\, e\, {\rm cm}\,\, .
\end{eqnarray}

Unfortunately, the various nuclear contributions partially cancel for the helion EDM, whereas they add up for the experimentally less interesting triton EDM. This cancellation is the origin of the rather large relative uncertainty of the total helion EDM multi-body contribution. This cancellation was found to be less profound in Ref.~\cite{Dekens:2014jka} since the $\Delta$-dependent correction to $g_1^\theta$ was not taken into account.
The uncertainties in Eq.~(\ref{triEDMtheta}) are dominated by the hadronic uncertainty of the coupling constant 
$g_1^\theta$, by
the nuclear and  hadronic uncertainties of the  $g_0^\theta$ term,\footnote{The hadronic uncertainties of $g_0^\theta$ and $g_1^\theta$  can be reduced by refined predictions of $c_1$, $\delta m_{np}^{\text{str}}$ and the quark mass ratio $m_u/m_d$ or difference $\epsilon$,
while the nuclear uncertainty might improve by the use of \NthreeLO\, CP-conserving chiral potentials, see the
discussion in Ref.~\cite{Dekens:2014jka}.} and, finally, by the intrinsic uncertainty due to  the CP-violating nucleon-nucleon contact interactions. 
The latter uncertainty, roughly $\pm 0.2\cdot 10^{-16}\, \bar\theta\cdot e\cdot {\rm cm}$, can be interpreted as the one arising from higher-order corrections and will be difficult to reduce.

\setcounter{equation}{0}
\section{The minimal left-right symmetric scenario}
\label{sec:LR}
In this section the implications of our results for the mLRSM scenario are briefly explained. This model and its induced hadronic coupling constants were discussed in detail in the context of hadronic EDMs in 
Ref.~\cite{Dekens:2014jka} (see also Refs.~\cite{Zhang:2007da,Maiezza:2014ala,Dekens:2014ina}). 
The predictions of the hadronic coupling constants of Ref.~\cite{Dekens:2014jka} as functions of $\Delta^{LR}$ are briefly summarized here before returning the focus on the implications of the results of the nuclear computations presented in this paper. 

The mLRSM is based on unbroken parity at high energies by extending the SM gauge symmetry 
to $SU(2)_R$~\cite{Pati:1974yy,Mohapatra:1974hk,Mohapatra:1974gc,Senjanovic:1975rk,Minkowski:1977sc,Senjanovic:1978ev,Mohapatra:1979ia,Mohapatra:1980yp}. 
Once the additional degrees of freedom -- in particular right-handed massive gauge bosons -- are integrated out at low energies, the effective Lagrangian contains an additional source of hadronic CP violation in the form of a particular four-quark operator, called the four-quark left-right (FQLR)  
operator~\cite{deVries:2012ab}.
The FQLR  operator not only breaks CP symmetry but also chiral and isospin symmetry non-trivially, resulting in a unique pattern of hadronic CP-violating interactions \cite{deVries:2012ab,dissertation}. 
We assume   that there  does not appear a $\theta$-term
in this model. For discussions of EDMs in left-right models with a nonzero $\theta$-term see, {\it e.g.}, Refs.~\cite{Maiezza:2014ala,Kuchimanchi:2014ota}.

 The FQLR operator induces the three-pion vertex in Eq.~\eqref{eq:impcoup} with coupling constant 
 $\Delta^{LR}$ as the leading term in the pion sector. According to Refs.~\cite{deVries:2012ab,dissertation},
 the leading contributions to the CP-violating pion-nucleon coupling constants are  then the following 
 functions of $\Delta^{LR}$:
  \begin{eqnarray}\label{g01LR}
   g_1^{LR}&=& 8 c_1 \mn \Delta^{LR} = (-7.5\pm 2.3) \Delta^{LR}\, \,,\nonumber\\
   g_0^{LR} &=& \frac{\delta m_{np}^{\rm str} \mn \Delta^{LR}} {M_\pi^2}  = (0.12\pm 0.02) \Delta^{LR} \,\,.
\end{eqnarray}
Independent contributions to $ g_{1,0}^{LR}$ appear at the same order, which scale as $\tilde g_1^{LR} = \mathcal{O}(\Delta^{LR})$ and $\tilde g_0^{LR} = \mathcal{O}( \epsilon \Delta^{LR} \mpi^2/\mn^2)\simeq 0.01\, \Delta^{LR}\,$ by NDA and are absorbed into the uncertainties of Eq.~\eqref{g01LR} here.
The main result is that the ratio ${g_0^{LR}}/{g_1^{LR}} \simeq -0.02\pm0.01$ is heavily suppressed, such that contributions to hadronic EDMs proportional to $g_0^{LR}$ can be neglected \cite{deVries:2012ab}.

Moreover, the coefficients $C_{1,2}$ of the isospin-symmetric nucleon-nucleon contact terms are heavily suppressed in the mLRSM due to the need of extra isospin violation. Therefore, these contact terms appear at \NfourLO\ and can be neglected. There are in principle contributions at \NtwoLO, one order higher than considered in this paper, from the isospin-breaking and 
CP-violating nucleon-nucleon contact terms
\begin{equation}
\mathcal L_{C_{3,4}} = C_3N^{\dagger}\tau_3N\,\mathcal{D}_{\mu}(N^{\dagger}S^{\mu}N)
                                   +C_4N^{\dagger}N\,\mathcal{D}_{\mu}(N^{\dagger}\tau_3\,S^{\mu}N)\,\,.
 \label{eq:Lc34}
\end{equation}
In analogy to the $\theta$-term case, contributions from these contact terms are regarded as the intrinsic uncertainties due to higher-order corrections. Their sizes can be assessed by considering two-pion-exchange diagrams induced by $g_1^{LR}$ to obtain $|C_{3,4}^{LR}|= \mathcal{O}(g_1^{LR} g_A/(\fpi \mn^2)) \simeq \Delta^{LR} \,\text{fm}^{3}$. Their contributions to the deuteron and three-nucleon EDMs read
\begin{eqnarray}
d^{LR}_{^2{\rm H}}(C_{3,4}^{LR}) &\simeq&  (0.05\pm0.05)\cdot \left(C_4^{LR}-C_3^{LR}\right)\,e \, \mathrm{fm}^{-2}\,\,,\label{eq:c34h2}\\
d^{LR}_{^3{\rm {He}}}(C_{3,4}^{LR})\simeq d^{LR}_{^3{\rm {H}}}(C_{3,4}^{LR})&\simeq&\left[(0.04\pm0.03)\cdot C_3^{LR}
-(0.07\pm 0.03)\cdot C_4^{LR}\right] e 
\, \mathrm{fm}^{-2}\,\,,\label{eq:c34he3h3}
\end{eqnarray}
based on the \NtwoLO\ $\chi$EFT potential.

\renewcommand{\thefootnote}{\fnsymbol{footnote}}
The total nuclear contribution to the deuteron EDM induced by the left-right-symmetric scenario is then given by\footnotemark[4]\footnotetext[4]{For the $D$-wave corrected
form of the $d_n$ and $d_p$ weights in Eq.\,(\ref{eq:deutLR}) see Appendix~\ref{app:erratum}.}
\renewcommand{\thefootnote}{\arabic{footnote}}
\begin{eqnarray}\label{dLR}
d^{LR}_{^2{\rm H}} - d_p^{LR} - d_n^{LR}  &=& \Delta^{LR} \, \left[(1.37 \pm 0.13\pm 0.41) + (0.75\pm 0.14)
\pm 0.1\right]  
\,e\, {\rm fm}\nn  \\
  &=& \Delta^{LR}\,(2.1 \pm 0.5) \,  e\, {\rm fm}\,\,,
  \label{eq:deutLR}
\end{eqnarray}
where the first contribution is the one of the $g_1^{LR}$-induced one-pion exchange, the second -- about $55\%$ of the
first one --
is its three-pion-induced one-loop correction $\Delta^{LR}f_{g_1}$ and the third is the uncertainty from the isospin-violating nucleon-nucleon contact terms. 
This result for the deuteron EDM allows for an extraction of the parameter $\Delta^{LR}$ from EDM measurements of the deuteron, proton and neutron. The nuclear contributions to the helion and triton EDMs provide a consistency check where again the uncertainties of the single-nucleon contributions can safely
be neglected:
\begin{eqnarray}
d_{{}^3{\rm He}}^{LR}-0.90\, d_n^{LR}+0.03\,d_p^{LR} &=& \Delta^{LR}\,\bigl\{ 
  (0.017\pm 0.006) -  (0.013\pm 0.002 \pm 0.002) 
  \nn \\
  &&\qquad+(1.07 \pm 0.14\pm 0.32)
+(0.61\pm 0.14) \pm 0.1 \bigr\}\, e\, {\rm fm} \nonumber\\
&=& \Delta^{LR}\, ( 1.7 \pm 0.5 \bigr)\, e\, {\rm fm}\,\,,\label{3HeLR}\\
d_{{}^3{\rm H}}^{LR} -0.92\,d_p^{LR} +0.03\,d_n^{LR}
&=& \Delta^{LR}\,\bigl\{ 
  (0.017\pm 0.006) + (0.013\pm 0.002 \pm 0.002) 
   \nn \\&&\qquad+(1.04 \pm 0.14\pm 0.31)
+(0.60\pm 0.14) \pm 0.1 \bigr\}\, e\, {\rm fm} \nn\\
&=& \Delta^{LR}\, ( 1.7\pm 0.5 \bigr)\, e\, {\rm fm}\label{3HLR}\,\,.
 \label{eq:hyLR}
\end{eqnarray}
The first term in brackets is the $\Delta^{LR}$-induced 3-nucleon contribution, the second and third one stem from the $g_0^{LR}$-
and $g_1^{LR}$-induced one-pion exchanges, respectively, the fourth  corresponds to the $\Delta^{LR} f_{g_1}$ vertex correction, and the fifth is again the uncertainty from the isospin-violating nucleon-nucleon contact terms.
As it is the case in Eq.~(\ref{eq:deutLR}), the first uncertainty is  always the nuclear one and the second, if displayed, is
the hadronic one.

The results of Ref.~\cite{Seng14} indicate that the single-nucleon EDMs induced by the FQLR operator are significantly smaller than the two- and three-nucleon contributions presented above, although there exists a considerable uncertainty. If the single-nucleon EDM contributions are neglected, there is a non-trivial relation between the considered light-nuclei EDMs induced by the mLRSM~\cite{Dekens:2014jka}:
\begin{equation}
d_{{}^3{\rm He}}^{LR}\simeq d_{{}^3{\rm H}}^{LR}\simeq 0.8\,\, d_{{}^2{\rm H}}^{LR}\,\,.
\end{equation}
Note especially that all contributions have the same sign in contradistinction to the $\theta$-term case.

\setcounter{equation}{0}
\section{Summary and conclusions}
\label{sec:summary}
In this work we have calculated the EDMs of the deuteron, helion, and triton in the framework of chiral effective field theory. The CP-conserving and -violating nucleon-nucleon potentials and currents are treated on an equal footing and were derived systematically within a controlled power-counting scheme. Up to next-to-leading order in the $\chi$EFT power-counting scheme, nuclear EDMs depend at most on the seven CP-violating hadronic interactions defined in 
Eq.~\eqref{eq:impcoup}, irrespectively of the underlying source of CP violation \cite{deVries2011b, Bsaisou:2012rg, deVries:2012ab,dissertation}. 

We have performed numerical calculations of the EDMs of the three lightest  nuclei as functions of these seven coupling constants. Wherever possible, we have compared our results with existing results in the 
literature~\cite{LiuTimmermans, Stetcu:2008vt,Afnan, deVries2011b,Song:2012yh} based on  phenomenological CP-conserving
potentials and found largely consistent results. While our results for the leading $g_{0}$- and $g_1$-induced EDM contributions are in agreement with those of Ref.~\cite{Song:2012yh}, they are smaller than those 
of Refs.~\cite{Stetcu:2008vt,deVries2011b} by a factor of two for the three-nucleon systems. Certain contributions, in particular those dependent on the CP-violating three-pion vertex, have been calculated in this work for the first time.
The consistent treatment within $\chi$EFT enabled us to compute the EDMs with well-defined {\it nuclear} uncertainties which arise from the cutoff dependence of the employed CP-conserving nuclear potential.
This uncertainty amounts to approximately $10\%$ for long-range contributions and to almost $50\%$ for short-range contributions. The main results of our work are given in Eqs.~\eqref{eq:h2edm}--\eqref{eq:h3edm}, which summarize the dependence of light-nuclei EDMs on the seven coupling constants. These results are model-independent, {\it i.e.} they are applicable to any model of CP violation.

In particular, which of the seven interactions dominate(s) the nuclear EDMs does
depend on the underlying mechanism of CP violation. However, one can still draw some general conclusions. First of all, contributions from CP-violating nucleon-nucleon contact interactions proved to be less suppressed with respect to one-pion-exchange contributions than 
chiral power-counting rules indicate. This observation increases the uncertainty of nuclear EDM calculations, but the extent depends on the underlying CP-violating source(s). In addition, this implies that calculations of CP-violating moments of heavier nuclei should not, in general, be performed on the basis of one-pion-exchange potentials only -- as is currently state of the art~\cite{Engel:2013lsa} -- in order to obtain reliable error estimates. This especially affects analyses of CP-violating models inducing a large gluon chromo-EDM which generates relatively large CP-violating nucleon-nucleon contact terms \cite{deVries2011b}.

Second, we find a significant contribution to nuclear EDMs arising from the one-loop correction of the $g_1$ vertex of Ref.~\cite{deVries2011b}, which is induced by the CP-violating three-pion vertex. The nuclear contributions from this correction turn out to be well approximated by their value at zero-momentum transfer. This means that the three-pion vertex effectively renormalizes the coupling constant $g_1$ in the case of light nuclei. However, the induced form factor grows linearly with the momentum transfer, which renders momentum-dependent corrections (that cannot be absorbed into $ g_1$) potentially more important for {\em heavier} systems as the Fermi-scale increases. A detailed calculation for such systems is necessary to quantify this effect. 

Third, the three-pion vertex induces a CP-violating three-nucleon potential which power-counting predicts to be significant. However, the full calculations performed here reveal that the three-body potential provides a negligible contribution to the EDMs of the considered three-nucleon systems. Symmetry or other constraints specific to the triton and helion wave functions can be excluded as the reason for this suppression.
This CP-violating three-nucleon potential might therefore be safely neglected in nuclear EDM calculations. 

Our EDM results can be used to investigate various specific scenarios of CP violation. As two examples, the QCD $\theta$-term and the minimal left-right symmetric scenario were considered here, which can both be traced back to only one dimensionless parameter -- of fundamental nature
in the former case and of only low-energy effective nature in the latter one. These parameters were discussed in detail in the context of EDMs in Ref.~\cite{Dekens:2014jka}.
The $\theta$-term scenario has the advantage that the coupling constants appearing in Eq.~\eqref{eq:impcoup} can be related to known strong matrix elements due to chiral-symmetry considerations. This led to predictions for the nuclear contributions to the deuteron, helion, and triton EDMs as functions of $\bar\theta$ directly, see Eqs.~\eqref{dEDMtheta} and \eqref{triEDMtheta}. The uncertainties of the deuteron and triton EDMs are quite small (roughly $30\%$ and $18\%$, respectively). Unfortunate cancellations among the various nuclear EDM contributions yield a somewhat larger uncertainty ($42\%$) for the experimentally interesting helion EDM.

The uncertainties of our results are governed
 by the nuclear uncertainty of the isospin-conserving CP-violating one-pion-exchange  and  nucleon-nucleon contact  terms, and especially, by the
 \textit{hadronic} uncertainties, which arise from the errors of the coupling constants of CP-violating operators. The hadronic uncertainties for the $\theta$-term scenario can be reduced with refined knowledge of the strong part of the neutron-proton mass splitting, $\delta m_{np}^{\mathrm{str}}$, and $c_1$ which is related to the pion-nucleon sigma term. Their uncertainties are expected to decrease with new Lattice-QCD predictions and a refined analysis of pion-nucleon scattering data (we refer to Ref.~\cite{Dekens:2014jka} for further details). In addition, improved Lattice-QCD calculations of the single-nucleon EDMs would allow for more precise predictions of the total nuclear EDMs. In this case two EDM measurements would be sufficient to confirm the existence of a nonzero $\theta$-term. The nuclear uncertainty of our results can be reduced by the application of \NthreeLO\,  chiral potentials and associated wave functions.

In the mLRSM scenario it is again possible to use chiral-symmetry considerations to greatly simplify the analysis. All nuclear contributions to the EDMs can be expressed as functions of a single coupling constant, see Eqs.~\eqref{dLR}--\eqref{3HLR}. Assuming the dominance of the nuclear EDM contributions over the single-nucleon EDM contributions, as expected from chiral perturbation theory \cite{Seng14}, the mLRSM predicts the deuteron, helion, and triton EDMs to be of the same sign and (approximately) magnitude. EDM measurements of single nucleons and light nuclei would thus be able to confirm/exclude the mLRSM as the primary origin of the measured EDMs. 

More general models can of course be studied in a similar fashion. However, the analysis is then limited by the unknown sizes of the various coupling constants appearing in the  model-independent EDM expressions, see Eqs.\,(\ref{eq:h2edm})--(\ref{eq:h3edm}). Nevertheless, general statements can still be made using estimates of these coupling constants with unfortunately larger uncertainties (see {\it e.g.} Refs.~\cite{deVries2011b, Dekens:2014jka}). For instance, if the hadronic CP violation is dominated by the quark EDM, the EDMs of light nuclei can be expressed in terms of the
 single-nucleon EDMs only. On the other hand, in models that induce a large gluon chromo-EDM almost all interactions in Eq.~\eqref{eq:impcoup} contribute to nuclear EDMs at the same order, which makes the analysis extremely complicated. The situation would improve substantially with Lattice-QCD calculations of the coupling constants of CP-violating effective Lagrangian terms induced by the various dimension-six CP-violating operators, see {\it e.g.} Refs.~\cite{Bhattacharya:2014cla,Shindler:2014oha}. 

In summary, we have performed calculations of light-nuclei EDMs in the framework of chiral effective field theory. We have included CP-violating one-, two-, and three-nucleon interactions up to next-to-leading order in the chiral power counting. We have shown that certain contributions to nuclear EDMs, {\it e.g.} from nucleon-nucleon contact interactions and the three-pion vertex, which are often neglected in the literature, are actually significant. As applications of our results, we have studied two specific scenarios of CP violation and demonstrated that these could be disentangled with EDM measurements of nucleons and light nuclei. We stress that an important and outstanding challenge in this field is the analog of our Eqs.~(\ref{eq:h2edm})--(\ref{eq:h3edm}) for heavier systems. 

\section*{Acknowledgements}
We are very grateful to Evgeny Epelbaum and Timo L\"ahde for useful
communications.
We would  also like to thank Stephan D\"urr, Emanuele Mereghetti, Tom Luu, Andrea Shindler and Frank Rathmann for discussions.
This work is supported in part by the DFG and the NSFC through funds 
provided to the Sino-German CRC 110 ``Symmetries and the Emergence of 
Structure in QCD''. The resources of the J{\"u}lich Supercomputing Center at the 
Forschungszentrum J{\"u}lich, namely the supercomputers JUQUEEN and JUROPA, have been instrumental in the computations reported here.

\appendix
\section{Power counting of CP-violating nuclear operators}
\label{app:pc}

\begin{table*}[th!]
\begin{center}
\caption{Power-counting estimates of the leading EDM contributions induced by the coefficients of CP-violating operators defined in Eqs.\,(\ref{eq:impcoup}), 
(\ref{eq:ncurrent})--(\ref{eq:3npot}) and (\ref{eq:Lc34}).
The second column displays the estimates for a general source of CP violation, while the third and fourth column show the power-counting estimates for the $\theta$-term and  the mLRSM scenario, respectively. Estimates which should be enhanced/suppressed
relatively to the power counting are explicitly marked by a star/diamond.
The expressions on the right sides of the third and fourth column indicate whether an operator yields a leading-order (\LO), a next-to-leading-order (\NLO), etc.\ contribution. 
\label{tab:pc}}

\smallskip
\begin{tabular}{c ||  c l | l c | l c }
label &\multicolumn{2}{c}{general source}&\multicolumn{2}{|c}{$\theta$-term}& \multicolumn{2}{|c}{mLRSM} \\
\hline
\hline
$d_{n,p}$ & $d_{n,p}$&$\phantom{\hspace{-0.3cm}\mn^2/(\fpi^4\mpi^3)}$&$\bar{\theta}\,e\,\mpi^2/\mn^3$&(\NLO)&$\Delta^{LR}\,e\,\fpi/\mn^2$& (\NthreeLO)\\
$\Delta$ & $\Delta$&$\hspace{-0.3cm}e\,\fpi/\mpi^2\ \diamond$& $\bar{\theta}\,e\,\fpi\mpi^2/\mn^4\ \diamond $ &(\NtwoLO)&$\Delta^{LR}\, e\,\fpi/\mpi^2\ \diamond$&(\NLO)\\
$g_0$   &$g_0$&$\hspace{-0.3cm}e\,\fpi/\mpi^2$&$\bar{\theta}\,e\,\fpi/\mn^2$& (\LO)&$\Delta^{LR}\,e\,\,\fpi/(\mpi \mn)$&(\NtwoLO) \\
$g_1$   &$g_1$&$\hspace{-0.3cm}e\,\fpi/\mpi^2$&$\bar{\theta}\, e\,\fpi \mpi/\mn^3$ &(\NLO)&$\Delta^{LR}\,e\,\mn\fpi/\mpi^3$& (\LO) \\
$\Delta\,f_{g_1}$ & $\Delta$&$\hspace{-0.3cm}e\, \fpi/(\mpi \mn)\,\ast$& $\bar{\theta}\,e\,\fpi\mpi^3/\mn^5\ \ast$ &(\NthreeLO)& $\Delta^{LR}\,e\,\fpi/(\mpi \mn)\,\ast$& (\NtwoLO)\\
$C_{1,2}$& $C_{1,2}$&$\hspace{-0.3cm}e\,\fpi^2$&$\bar{\theta}\,e\,\fpi\mpi^2/\mn^4$ &(\NtwoLO)&$\Delta^{LR}\,e\,\fpi\mpi/\mn^3$& (\NfourLO)\\
$C_{3,4}$& $C_{3,4}$&$\hspace{-0.3cm}e\,\fpi^2$&$\bar{\theta}\,e\,\fpi\mpi^3/\mn^5$ &(\NthreeLO)&$\Delta^{LR}\,e\,\fpi /(\mpi \mn)$& (\NtwoLO)
\end{tabular}
\end{center}
\end{table*}
We estimate the contributions of CP-violating nuclear operators, 
defined by Eqs.\,(\ref{eq:impcoup}),
(\ref{eq:ncurrent})--(\ref{eq:3npot}) and (\ref{eq:Lc34}), 
 to the EDMs of  light nuclei. 
 The  power-counting scheme of Ref.~\cite{Liebig:2010ki} is employed, which is also used in 
\mbox{Refs.~\cite{Bsaisou:2012rg,dissertation}}\footnote{The operators relevant for this work are counted in the same way
as in the power-counting scheme of \mbox{Refs.~\cite{deVries2011b,deVries:2012ab}}. Differences between these schemes only emerge at one-loop level and
in certain currents which have been pointed out in Ref.~\cite{Bsaisou:2012rg}.}.
Within this scheme, the counting orders increase as integer powers of $p/\Lambda_{\chi}\sim\mpi/\mn$. 
To obtain the estimates we take out a common  factor from the corresponding diagrams -- for the three-body case 
 this normalization factor is $\mn^2/(\fpi^4 \mpi^3)$.\footnote{One factor of $\mpi$ for the photon momentum, {\it cf.}  Eq.~(\ref{eq:FFdef}), and one
 factor of $\fpi^{-2} \times \mn/\mpi^2$ per  one-pion exchange  extracted from the
nuclear wave function in  such a way that Fig.\,\ref{Fig:current}\,(a) becomes simply connected.}

 While the second column of Table~\ref{tab:pc}  contains the power-counting estimates for a general source of CP-violation
 with input according to  Eqs.\,(\ref{eq:ncurrent})--(\ref{eq:3npot}), 
 the third and  fourth columns list the estimates for  the $\theta$-term and the mLRSM scenarios,
respectively. For that purpose,
the results of Section~\ref{sec:theta} for the $\theta$-term case and of Section~\ref{sec:LR} for the mLRSM scenario
have been utilized to assess 
the numerical sizes of the coefficients, which do not follow in all cases NDA:
\begin{itemize}
\item
 $ |\gzero| \sim \bar{\theta} \cdot (\mpi/\mn)^2$, $|\gone|\sim(\mpi/\mn)\cdot |\gzero|$ and
$|\Delta^{\theta}| \sim (\mpi/\mn)^2 \cdot |\gzero|$ in the $\theta$-term case,
\item
$|g_1^{LR}|\sim\Delta^{LR}\cdot (\mpi/\mn)^{-1}$ and  $|g_0^{LR}|\sim\Delta^{LR}\cdot(  \mpi/\mn)$ in the mLRSM scenario,
and
\item
$|C_{1,2}|\sim |g_0|/(\fpi \mn^2)$ and $|C_{3,4}|\sim |g_1|/(\fpi \mn^2)$ in general.
\end{itemize}
The second row of Table~\ref{tab:pc}  is specific to a three-body potential and therefore does not apply to the  deuteron -- on top of the fact that  the third and sixth rows   are ruled out by isospin selection in this case.
The estimates marked by a star~($\ast$)  in Table~\ref{tab:pc}
 should be enhanced by a factor of about $5\pi$ relative to the stated order,
while the entries marked by a diamond ($\diamond$)  are in reality suppressed by  about
one order of $\mpi/\mn$  -- see Section~\ref{sec:nucedms} for more details. 


\section{Regulator dependence of the contact-interaction EDM terms}
\label{app:beta}

\begin{figure}[t]
\centering
\includegraphics[scale=0.6]{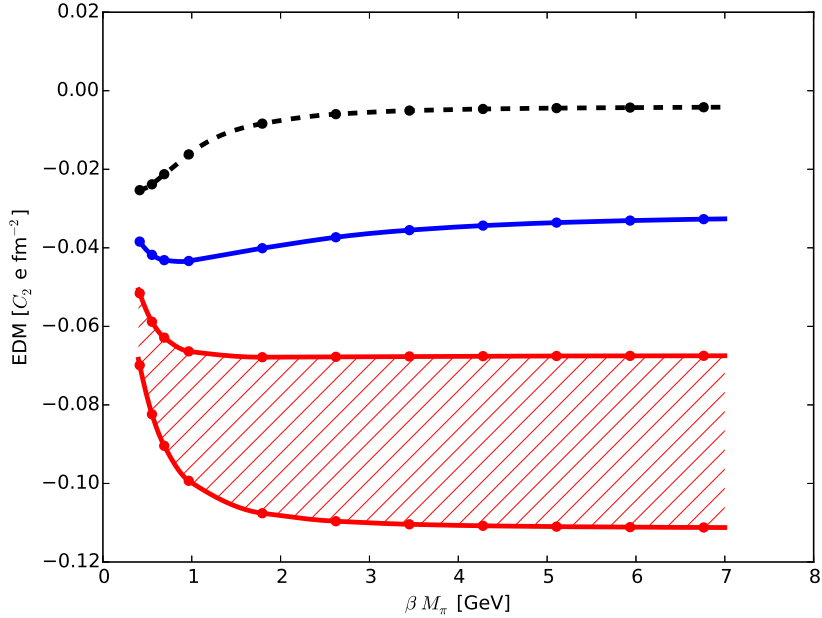}
\caption{Dependence of the helion EDM contribution induced by the $C_2$ vertex on the cutoff parameter $\beta$. The dashed line depicts the $\beta$ dependence when the A$v_{18}$+UIX potential \cite{av18,Pudliner:1997ck} is employed for the CP-conserving component of the nuclear potential, while the solid line shows the $\beta$ dependence for the CD-Bonn+TM potential \cite{cdbonn,Coon:2001pv}. The hatched area depicts the dependence of the $C_2$-induced EDM contribution on $\beta$ and on the cutoffs $\Lambda_{{\rm LS}}$ and $\Lambda_{{\rm SFR}}$ for the \NtwoLO\ $\chi$EFT potential \cite{Epelbaum:2004fk,chiralpotentials}.\label{fig:betadep}}
\end{figure}

In order to investigate the EDM contributions from the two-nucleon contact interactions in Eq.\,(\ref{eq:impcoup}) 
and Eq.\,(\ref{eq:Lc34}), an additional cutoff function with parameter $\beta$ has been introduced --- see 
the third line of Eq.\,(\ref{eq:nnpot}). As a study case, Figure \ref{fig:betadep} depicts the $\beta$ dependence of the contributions to the helion EDM induced by the $C_2$ vertex when  the 
CP-conserving component of the nuclear potential is  given, respectively, by the 
A$v_{18}$+UIX potential \cite{av18,Pudliner:1997ck}, the CD-Bonn+TM potential \cite{cdbonn,Coon:2001pv} or the \NtwoLO\ $\chi$EFT potential \cite{Epelbaum:2004fk,chiralpotentials} --- the latter  with the  five combinations of cutoffs as  in Eq.\,(\ref{eq:cutoffs}).  

Modulo a prefactor, 
the potential operator induced by the $C_2$ vertex in the third line of Eq.\,(\ref{eq:nnpot}) coincides with the $g_0$-induced potential operator in the first line of Eq.\,(\ref{eq:nnpot}) for $\beta=1$. The $g_0$-induced contributions to the 
helion EDM as listed in Table \ref{tab:3n} can thus be recovered at $\beta=1$ by a suitable replacement of units. We have verified explicitly that our numerical 
calculations are in agreement with this expectation.

The $C_i$ vertices parameterize physics at the momentum scale \mbox{$\gtrsim 3\mpi$}. 
For \mbox{$\beta\leq 3$}, the EDM contributions from the A$v_{18}$+UIX, the CD-Bonn+TM and the \NtwoLO\ $\chi$EFT potential with the five  cutoff combinations are  compatible within one order in magnitude since they only differ by a factor of less than three. 

To $1\%$ accuracy, the EDM contributions from the \NtwoLO\ $\chi$EFT potential  have converged  already if
$\beta\mpi > 3\,{\rm GeV}$.
The convergence of the corresponding  EDM contributions from the A$v_{18}$+UIX and the CD-Bonn+TM potentials, however, is 
more slowly. The discrepancies between the EDM contributions from the three different CP-conserving potentials are especially significant at large $\beta\mpi$. This reveals the tremendous model dependence in the short-distance regime. The large $\beta$ limit of A$v_{18}$+UIX differs from the one of CD-Bonn+TM by a factor of about eight. 
As already discussed in Section~\ref{sec:nucedms}, the small value of the 
A$v_{18}$+UIX  limit can be attributed to a large atypical short-range repulsion. The absolute distance between the large $\beta$ limit of  the CD-Bonn+TM case  and the $\beta$  band of the \NtwoLO\ $\chi$EFT potential is roughly the same
as between the $\beta$ limits of the  CD-Bonn+TM and A$v_{18}$+UIX cases. The values at \mbox{$\beta=49$} were taken as the predictions for the short range EDM contributions given in the last four rows of Table \ref{tab:3n} and Eqs.\,(\ref{eq:c34h2})-(\ref{eq:c34he3h3}). 
 
 The patterns of convergence with respect to $\beta$ of the helion EDM contributions induced by the other $C_i$ vertices as well as of their corresponding triton counter parts are similar. Thus they are not explicitly shown here.

\section{Erratum  after publication}  \label{app:erratum}
As first observed in  Ref.~\cite{Yamanaka:2015qfa},
the usual weight factors of the neutron  ($d_n$) and proton  ($d_p$)  single-nucleon contribution
to the electric dipole moment of the deuteron 
are lacking a small wave-function-dependent
 term resulting from  the subleading ${}^3\!D_1$ component of the deuteron wave function.
A simple calculation reveals that the total single-nucleon contribution is 
\[
d_{^2{\rm H},{\rm single}} = \left(1  -\frac{3}{2} \, P_{D} \right) \, (d_n + d_p ) \,,
\]
where $P_{D}$ is the probability of the deuteron ${}^3\!D_1$-state, which of course depends on
the choice of the wave function.

\medskip
Therefore  the 
values $1.00$  of the $d_n$ and $d_p$ weight factors  have to be modified in the following places
of this paper:
\begin{enumerate}
\item[(i)] in the first two rows of Table~\ref{tab:deuteron},  
see the enclosed Table~\ref{tab:deuteron_prime},
\item[(ii)] in the first bracket on the right-hand side of Eq.\,(\ref{eq:h2edm}), 
\[
d_{^2{\rm H}} =
 {(0.939\pm0.009)(d_n + d_p )}
- \bigl [ (0.183 \pm 0.017) \,g_1
- (0.748\pm 0.138) \,\Delta\bigr] \,e \,{\rm fm} \,,
\tag{\mbox{\ref{eq:h2edm}}$^\prime$}
\]
\item[(iii)] implicitly in the first bracket of Eq.\,(\ref{eq:d2Htheta}),
\[
d^\theta_{{}^2{\rm H}} =
   \bar\theta\cdot \left\{  \bigl[(0.56 \pm 0.01 \pm 1.59)\bigr] 
  - (0.62 \pm 0.06\pm 0.28) - (0.28\pm 0.05\pm 0.07)
   \right\}\cdot 10^{-16}\,  e\, {\rm cm}\,,
    \tag{\mbox{\ref{eq:d2Htheta}}$^\prime$}
\]
such that ``$(0.6\pm 1.7)$" is replaced by ``$(0.56 \pm 0.01 \pm 1.59)$",
where the first uncertainty is the nuclear one, while the second is the
hadronic one, 
\item[(iv)]  
explicitly on the left-hand side of Eq.\,(\ref{dEDMtheta}), 
\[
d^\theta_{{}^2{\rm H}} - 0.94(d_p^\theta +d_n^\theta) =- \bar\theta\cdot (0.89 \pm 0.30) \cdot 10^{-16}\,  e\, {\rm cm}\,\,,
\tag{\mbox{\ref{dEDMtheta}}$^\prime$}
\]
\item[(v)] and on the left-hand side of Eq.\,(\ref{eq:deutLR}),
\[
\begin{array}{lcl}
d^{LR}_{^2{\rm H}} -  0.94(d_p^{LR} +d_n^{LR} ) &=& \Delta^{LR} \, \left[(1.37 \pm 0.13\pm 0.41) + (0.75\pm 0.14)
\pm 0.1\right]  
\,e\, {\rm fm}  \nonumber \\
  &=& \Delta^{LR}\,(2.1 \pm 0.5) \,  e\, {\rm fm}\,\,.
  \end{array} \tag{\mbox{\ref{eq:deutLR}}$^\prime$}
\]
\end{enumerate}
In the latter two cases the uncertainty of the single-nucleon contributions can safely be neglected.

\setcounter{table}{0}
\renewcommand{\thetable}{\arabic{table}$^\prime$}
\begin{table*}[t]
\begin{center}
\caption{The new entries of Table~1. Captions as in Table~1. Note that  the new   $d_n$ and $d_{p}$ weight factors displayed for the A$v_{18}$ potential exactly agree with the 
ones of Ref.~\cite{Yamanaka:2015qfa}.  
\label{tab:deuteron_prime}}

\smallskip
\begin{tabular}{c || c | c | c || c}
label & \NtwoLO\ $\chi$EFT& A$v_{18}$  & CD-Bonn & units \\
\hline
\hline
$d_{n}$               & $\phantom{-}0.939 \pm 0.009$ & $\phantom{-}0.914$ & $\phantom{-}0.927$ & $d_n$\\
$d_{p}$               & $\phantom{-}0.939 \pm 0.009$ & $\phantom{-}0.914$ & $\phantom{-}0.927$ & $d_p$\\
$g_1$                  & $ -0.183\pm 0.017 $                 & $-0.186$                  & $-0.186$ & $g_1\,e\,{\rm fm}$ \\
$\Delta\,f_{g_1}$ & $\phantom{-}0.748\pm0.138$   &$\phantom{-}0.703$ &$\phantom{-}0.719$ & $\Delta \,e\,{\rm fm}$
\end{tabular}
\end{center}
\end{table*}


\bibliographystyle{h-physrev5} 
 \bibliography{bibliography}

\begin{thebibliography}{10}

\bibitem{Czarnecki:1997bu}
A.~Czarnecki and B.~Krause,
\newblock Phys. Rev. Lett. {\bf 78}, 4339 (1997), arXiv:hep-ph/9704355.

\bibitem{Pospelov_review}
M.~Pospelov and A.~Ritz,
\newblock Annals Phys. {\bf 318}, 119 (2005), arXiv:hep-ph/0504231.

\bibitem{Mannel:2012qk}
T.~Mannel and N.~Uraltsev,
\newblock Phys. Rev. D {\bf 85}, 096002 (2012), arXiv:1202.6270.

\bibitem{Mannel:2012hb}
T.~Mannel and N.~Uraltsev,
\newblock JHEP {\bf 03}, 064 (2013), arXiv:1205.0233.

\bibitem{'tHooft:1976up}
G.~'t~Hooft,
\newblock Phys. Rev. Lett. {\bf 37}, 8 (1976).

\bibitem{Buchmuller:1982ye}
W.~Buchm{\"u}ller and D.~Wyler,
\newblock Phys. Lett. {\bf 121B}, 321 (1983).

\bibitem{Grzadkowski:2010es}
B.~Grzadkowski, M.~Iskrzynski, M.~Misiak, and J.~Rosiek,
\newblock JHEP {\bf 10}, 085 (2010), arXiv:1008.4884.

\bibitem{deVries:2012ab}
J.~de~Vries, E.~Mereghetti, R.~G.~E. Timmermans, and U.~van Kolck,
\newblock Annals Phys. {\bf 338}, 50 (2013), arXiv:1212.0990.

\bibitem{khriplovich}
I.~B. Khriplovich and R.~A. Korkin,
\newblock Nucl. Phys. A {\bf 665}, 365 (2000), arXiv:nucl-th/9904081.

\bibitem{Pospelov_deuteron}
O.~Lebedev, K.~A. Olive, M.~Pospelov, and A.~Ritz,
\newblock Phys. Rev. D {\bf 70}, 016003 (2004), arXiv:hep-ph/0402023.

\bibitem{LiuTimmermans}
C.-P. Liu and R.~G.~E. Timmermans,
\newblock Phys. Rev. C {\bf 70}, 055501 (2004), arXiv:nucl-th/0408060.

\bibitem{Stetcu:2008vt}
I.~Stetcu, C.-P. Liu, J.~L. Friar, A.~C. Hayes, and P.~Navratil,
\newblock Phys. Lett. B {\bf 665}, 168 (2008), arXiv:0804.3815.

\bibitem{Afnan}
I.~R. Afnan and B.~F. Gibson,
\newblock Phys. Rev. C {\bf 82}, 064002 (2010), arXiv:1011.4968.

\bibitem{deVries2011a}
J.~de~Vries, E.~Mereghetti, R.~G.~E. Timmermans, and U.~van Kolck,
\newblock Phys. Rev. Lett. {\bf 107}, 091804 (2011), arXiv:1102.4068.

\bibitem{deVries2011b}
J.~de~Vries {\em et~al.},
\newblock Phys. Rev. C {\bf 84}, 065501 (2011), arXiv:1109.3604.

\bibitem{Bsaisou:2012rg}
J.~Bsaisou {\em et~al.},
\newblock Eur. Phys. J. A {\bf 49}, 31 (2013), arXiv:1209.6306.

\bibitem{Song:2012yh}
Y.-H. Song, R.~Lazauskas, and V.~Gudkov,
\newblock Phys. Rev. C {\bf 87}, 015501 (2013), arXiv:1211.3762.

\bibitem{dissertation}
J.~Bsaisou,
\newblock {\em {Electric Dipole Moments of Light Nuclei}},
\newblock Dissertation, University of Bonn, 2014.

\bibitem{Dekens:2014jka}
W.~Dekens {\em et~al.},
\newblock JHEP {\bf 07}, 069 (2014), arXiv:1404.6082.

\bibitem{Wirzba:2014mka}
A.~Wirzba,
\newblock Nucl. Phys. A {\bf 928}, 116 (2014), arXiv:1404.6131.

\bibitem{Semertzidis:2003iq}
EDM Collaboration, Y.~Semertzidis {\em et~al.},
\newblock AIP Conf. Proc. {\bf 698}, 200 (2004), arXiv:hep-ex/0308063.

\bibitem{Semertzidis:2011qv}
Storage Ring EDM Collaboration, Y.~K. Semertzidis,
\newblock (2011), arXiv:1110.3378.

\bibitem{Lehrach}
A.~Lehrach, B.~Lorentz, W.~Morse, N.~Nikolaev, and F.~Rathmann,
\newblock (2012), arXiv:1201.5773.

\bibitem{Pretz:2013us}
J.~Pretz,
\newblock Hyperfine Interact. {\bf 214}, 111 (2013), arXiv:1301.2937.

\bibitem{Rathmann:2013rqa}
JEDI and srEDM Collaborations, F.~Rathmann, A.~Saleev, and N.~N. Nikolaev,
\newblock J. Phys. Conf. Ser. {\bf 447}, 012011 (2013).

\bibitem{Liebig:2010ki}
S.~Liebig, V.~Baru, F.~Ballout, C.~Hanhart, and A.~Nogga,
\newblock Eur. Phys. J. A {\bf 47}, 69 (2011), arXiv:1003.3826.

\bibitem{pdg}
Particle Data Group, K.~A. Olive {\em et~al.},
\newblock Chin. Phys. C {\bf 38}, 090001 (2014).

\bibitem{Bernard:1995dp}
V.~Bernard, N.~Kaiser, and U.-G. Mei{\ss}ner,
\newblock Int. J. Mod. Phys. E {\bf 4}, 193 (1995), arXiv:hep-ph/9501384.

\bibitem{Maekawa:2011vs}
C.~M. Maekawa, E.~Mereghetti, J.~de~Vries, and U.~van Kolck,
\newblock Nucl. Phys. A {\bf 872}, 117 (2011), arXiv:1106.6119.

\bibitem{Bernard:1991rt}
V.~Bernard, N.~Kaiser, J.~Gasser, and U.-G. Mei{\ss}ner,
\newblock Phys. Lett. B {\bf 268}, 291 (1991).

\bibitem{Becher:1999he}
T.~Becher and H.~Leutwyler,
\newblock Eur. Phys. J. C {\bf 9}, 643 (1999), arXiv:hep-ph/9901384.

\bibitem{Friar:2003yv}
J.~L. Friar, U.~van Kolck, G.~L. Payne, and S.~A. Coon,
\newblock Phys. Rev. C {\bf 68}, 024003 (2003), arXiv:nucl-th/0303058.

\bibitem{Baru:2012iv}
V.~Baru {\em et~al.},
\newblock Eur. Phys. J. A {\bf 48}, 69 (2012), arXiv:1202.0208.

\bibitem{Epelbaum:2004fk}
E.~Epelbaum, W.~Gl{\"o}ckle, and U.-G. Mei{\ss}ner,
\newblock Nucl. Phys. A {\bf 747}, 362 (2005), arXiv:nucl-th/0405048.

\bibitem{chiralpotentials}
E.~Epelbaum, H.-W. Hammer, and U.-G. Mei{\ss}ner,
\newblock Rev. Mod. Phys. {\bf 81}, 1773 (2009), arXiv:0811.1338.

\bibitem{Valderrama:2011hz}
M.~P. Valderrama,
\newblock Phys. Rev. C {\bf 83}, 024003 (2011).

\bibitem{Valderrama:2011hw}
M.~P. Valderrama,
\newblock Phys. Rev. C {\bf 84}, 064002 (2011).

\bibitem{av18}
R.~B. Wiringa, V.~G.~J. Stoks, and R.~Schiavilla,
\newblock Phys. Rev. C {\bf 51}, 38 (1995), arXiv:nucl-th/9408016.

\bibitem{Pudliner:1997ck}
B.~S. Pudliner, V.~R. Pandharipande, J.~Carlson, S.~C. Pieper, and R.~B.
  Wiringa,
\newblock Phys. Rev. C {\bf 56}, 1720 (1997), arXiv:nucl-th/9705009.

\bibitem{cdbonn}
R.~Machleidt,
\newblock Phys. Rev. C {\bf 63}, 024001 (2001), arXiv:nucl-th/0006014.

\bibitem{Coon:2001pv}
S.~A. Coon and H.~K. Han,
\newblock Few Body Syst. {\bf 30}, 131 (2001), arXiv:nucl-th/0101003.

\bibitem{Yamanaka:2014nba}
N.~Yamanaka, T.~Sato, and T.~Kubota,
\newblock JHEP {\bf 12}, 110 (2014), arXiv:1406.3713.

\bibitem{BiraEmanuele}
E.~Mereghetti, W.~H. Hockings, and U.~van Kolck,
\newblock Annals Phys. {\bf 325}, 2363 (2010), arXiv:1002.2391.

\bibitem{Aoki:2013ldr}
S.~Aoki {\em et~al.},
\newblock Eur. Phys. J. C {\bf 74}, 2890 (2014), arXiv:1310.8555v4.

\bibitem{Baru:2011bw}
V.~Baru {\em et~al.},
\newblock Nucl. Phys. A {\bf 872}, 69 (2011), arXiv:1107.5509.

\bibitem{Walker-Loud:2014iea}
A.~Walker-Loud,
\newblock PoS {\bf LATTICE\,2013}, 013 (2014), arXiv:1401.8259.

\bibitem{Borsanyi:2014jba}
S.~Borsanyi {\em et~al.},
\newblock Science {\bf 347}, 1452 (2015), arXiv:1406.4088.

\bibitem{Guo12}
F.-K. Guo and U.-G. Mei{\ss}ner,
\newblock JHEP {\bf 12}, 097 (2012), arXiv:1210.5887.

\bibitem{Akan:2014yha}
T.~Akan, F.-K. Guo, and U.-G. Mei{\ss}ner,
\newblock Phys. Lett. B {\bf 736}, 163 (2014), arXiv:1406.2882.

\bibitem{Shintani:2008nt}
E.~Shintani, S.~Aoki, and Y.~Kuramashi,
\newblock Phys. Rev. D {\bf 78}, 014503 (2008), arXiv:0803.0797.

\bibitem{Shintani:2012zca}
E.~Shintani, T.~Blum, and T.~Izubuchi,
\newblock PoS {\bf Confinement\,X}, 330 (2012).

\bibitem{Shintani:2014}
E.~Shintani,
\newblock {Lattice calculation of nucleon EDM},
\newblock { Talk given at ``Hadrons from Quarks and Gluons'', Hirschegg,
  Austria}, {2014}.

\bibitem{Filin:2009yh}
A.~Filin {\em et~al.},
\newblock Phys. Lett. B {\bf 681}, 423 (2009), arXiv:0907.4671.

\bibitem{Baru:2013zpa}
V.~Baru, C.~Hanhart, and F.~Myhrer,
\newblock Int. J. Mod. Phys. E {\bf 23}, 1430004 (2014), arXiv:1310.3505.

\bibitem{Adlarson:2014yla}
WASA-at-COSY Collaboration, P.~Adlarson {\em et~al.},
\newblock Phys. Lett. B {\bf 739}, 44 (2014), arXiv:1407.2756.

\bibitem{Ottnad:2009jw}
K.~Ottnad, B.~Kubis, U.-G. Mei{\ss}ner, and F.-K. Guo,
\newblock Phys. Lett. B {\bf 687}, 42 (2010), arXiv:0911.3981.

\bibitem{Mer11}
E.~Mereghetti, J.~de~Vries, W.~H. Hockings, C.~M. Maekawa, and U.~van Kolck,
\newblock Phys. Lett. B {\bf 696}, 97 (2011), arXiv:1010.4078.

\bibitem{CDVW79}
R.~J. Crewther, P.~Di~Vecchia, G.~Veneziano, and E.~Witten,
\newblock Phys. Lett. {\bf 88B}, 123 (1979),
\newblock [{\em Erratum ibid.} {\bf 91B}, 487 (1980)].

\bibitem{Zhang:2007da}
Y.~Zhang, H.~An, X.~Ji, and R.~N. Mohapatra,
\newblock Nucl. Phys. B {\bf 802}, 247 (2008), arXiv:0712.4218.

\bibitem{Maiezza:2014ala}
A.~Maiezza and M.~Nemevšek,
\newblock Phys. Rev. D {\bf 90}, 095002 (2014), arXiv:1407.3678.

\bibitem{Dekens:2014ina}
W.~Dekens and D.~Boer,
\newblock Nucl. Phys. B {\bf 889}, 727 (2014), arXiv:1409.4052.

\bibitem{Pati:1974yy}
J.~C. Pati and A.~Salam,
\newblock Phys. Rev. D {\bf 10}, 275 (1974),
\newblock [{\em Erratum ibid.} D {\bf 11}, 703 (1975)].

\bibitem{Mohapatra:1974hk}
R.~N. Mohapatra and J.~C. Pati,
\newblock Phys. Rev. D {\bf 11}, 566 (1975).

\bibitem{Mohapatra:1974gc}
R.~N. Mohapatra and J.~C. Pati,
\newblock Phys. Rev. D {\bf 11}, 2558 (1975).

\bibitem{Senjanovic:1975rk}
G.~Senjanovic and R.~N. Mohapatra,
\newblock Phys. Rev. D {\bf 12}, 1502 (1975).

\bibitem{Minkowski:1977sc}
P.~Minkowski,
\newblock Phys. Lett. {\bf 67B}, 421 (1977).

\bibitem{Senjanovic:1978ev}
G.~Senjanovic,
\newblock Nucl. Phys. B {\bf 153}, 334 (1979).

\bibitem{Mohapatra:1979ia}
R.~N. Mohapatra and G.~Senjanovic,
\newblock Phys. Rev. Lett. {\bf 44}, 912 (1980).

\bibitem{Mohapatra:1980yp}
R.~N. Mohapatra and G.~Senjanovic,
\newblock Phys. Rev. D {\bf 23}, 165 (1981).

\bibitem{Kuchimanchi:2014ota}
R.~Kuchimanchi,
\newblock Phys. Rev. D {\bf 91}, 071901 (2015), arXiv:1408.6382.

\bibitem{Seng14}
C.-Y. Seng, J.~de~Vries, E.~Mereghetti, H.~H. Patel, and M.~Ramsey-Musolf,
\newblock Phys. Lett. B {\bf 736}, 147 (2014), arXiv:1401.5366.

\bibitem{Engel:2013lsa}
J.~Engel, M.~J. Ramsey-Musolf, and U.~van Kolck,
\newblock Prog. Part. Nucl. Phys. {\bf 71}, 21 (2013), arXiv:1303.2371.

\bibitem{Bhattacharya:2014cla}
T.~Bhattacharya, V.~Cirigliano, and R.~Gupta,
\newblock PoS {\bf LATTICE\,2013}, 299 (2014), arXiv:1403.2445.

\bibitem{Shindler:2014oha}
A.~Shindler, J.~de~Vries, and T.~Luu,
\newblock PoS {\bf LATTICE\,2014}, 251 (2014), arXiv:1409.2735.

\bibitem{Yamanaka:2015qfa}
N.~Yamanaka and E.~Hiyama,
\newblock (2015), arXiv:1503.04446.

\end{thebibliography}

\end{document}